\documentclass[a4paper,11pt]{article}

\usepackage[svgnames]{xcolor}
\usepackage{jheppub}

\usepackage{graphicx}
\usepackage{subcaption}

\usepackage{float}
\usepackage{lipsum}
\usepackage{booktabs}
\usepackage{tabularx}
\usepackage{tikz}
\usepackage{pgfplots}
\usetikzlibrary{positioning,shapes,trees,arrows,graphs,decorations,calc}
\usepackage{url}

\usepackage[inline]{enumitem}

\usepackage{bm}
\usepackage{braket}
\newcommand{\betat}{\beta_t}

\usepackage{amssymb}
\usepackage{amsmath}
\usepackage{hyperref}

\usepackage{placeins} 
\usepackage[normalem]{ulem}
\usepackage{xspace}

\newcommand{\abs}[1]{\lvert#1\rvert}

\newcommand{\df}{\mathrm{d}}

\newcommand{\Tau}{\mathcal{T}}

\newcommand{\nn}{\nonumber}

\newcommand{\de}{\mathrm{d}}

\newcommand{\w}{\mathbf{w}}
\newcommand{\CF}{C_F}

\newcommand{\dR}{d_R}
\newcommand{\mat}[1]{\begin{pmatrix}#1\end{pmatrix}}

\newcommand{\tr}{\mathrm{tri}}

\pgfplotsset{compat=1.18}

\title{NNLO soft functions for heavy-quark pair production at hadron colliders}

\author[a]{Guido Bell,}
\emailAdd{bell@physik.uni-siegen.de}

\author[b]{Alessandro Broggio,}
\emailAdd{alessandro.broggio@univie.ac.at}

\author[a]{Sebastian Edelmann,}
\emailAdd{Sebastian.Edelmann@physik.uni-siegen.de}

\author[c]{Matthew A. Lim,}
\emailAdd{matthew.a.lim577@gmail.com}

\author[b]{and Rudi Rahn}
\emailAdd{rudi.rahn@univie.ac.at}

\affiliation[a]{Theoretische Physik 1, Center for Particle Physics Siegen, Universit\"at Siegen, Walter-Flex-Strasse 3, 57068 Siegen, Germany}
\affiliation[b]{Faculty of Physics, University of Vienna, Boltzmanngasse 5, A-1090 Wien, Austria}
\affiliation[c]{Department of Physics and Astronomy, University of Sussex, Sussex House, Brighton, BN1 9RH, UK}
\abstract{We generalise the {\tt SoftSERVE} framework for the automated computation of soft functions involving final-state heavy quarks in back-to-back kinematics up to next-to-next-to-leading order (NNLO) in the strong-coupling expansion. Our algorithm employs suitable phase-space parameterisations for both massive and mixed massless-massive contributions, allowing for an efficient numerical evaluation. Focusing on SCET$_\mathrm{I}$ observables that satisfy non-Abelian exponentiation, we present explicit results for the \mbox{0-jettiness} soft function in hadronic top-quark pair production in the form of two-dimensional grids. We describe the renormalisation procedure in detail and discuss the singular behaviour of the tripole colour correlations in the threshold limit.
}

\begin{document}

\preprint{%
\begin{tabular}[t]{r}
UWThPh 2026-7 \\
SI-HEP-2026-17 \\
P3H-26-061
\end{tabular}}

\begin{flushright}
\end{flushright}

\maketitle

\section{Introduction}

The High-Luminosity Large Hadron Collider (HL-LHC) is expected to begin operations before the end of the decade, providing the experiments with data samples that are an order of magnitude larger than those collected during the current LHC program. This unprecedented dataset will enable highly precise measurements across a broad range of processes and observables. In order to take full advantage of these novel opportunities, equally precise Standard Model (SM) predictions are needed, as any evidence for new physics at the HL-LHC is expected to manifest itself through subtle deviations from the SM predictions.

In light of these developments, precision calculations in QCD -- either based on fixed-order or resummed perturbation theory -- are becoming increasingly important. While fixed-order calculations rely on advanced techniques for the evaluation of multi-loop integrals, as well as efficient methods for the subtraction of infrared (IR) singularities, resummed predictions require a thorough understanding of the soft and collinear dynamics to all orders in perturbation theory. Matching these predictions to parton showers in Monte-Carlo event generators ultimately provides a realistic description of fully exclusive collider events.

Over the past decades, Soft-Collinear Effective Theory (SCET) has emerged as an efficient framework for precision calculations in perturbative QCD. In particular, SCET provides a systematic approach to deriving factorisation theorems and enables the use of renormalisation-group (RG) methods for the resummation of large logarithms. Within this framework, soft functions -- defined as correlators of Wilson lines -- are essential ingredients that describe the low-energy component of QCD radiation in the collider environment. They are required for achieving higher-order resummation, developing slicing and non-local subtraction techniques, and constructing Monte-Carlo event generators based on jet-resolution variables.

Progress in the computation of soft functions at next-to-next-to-leading order (NNLO) and beyond has, however, so far been largely restricted to processes involving only massless partons in the underlying hard scattering. This limitation does not stem from the absence of suitable factorisation theorems for more general processes and observables, but rather from the rapidly increasing computational complexity associated with massive partons, which renders fully analytic calculations exceedingly challenging.

Despite these challenges, a number of soft functions involving heavy quarks have been computed at NNLO in recent years. These include the threshold soft function
for top-pair production at lepton~\cite{vonManteuffel:2014mva} and hadron colliders~\cite{Wang:2018vgu} as well as for single-top production~\cite{Ding:2025xhc}, the small-transverse-momentum soft function for top-pair~\cite{Angeles-Martinez:2018mqh,Catani:2023tby} and associated top-pair production~\cite{Devoto:2025eyc}, the 0-jettiness soft function for single-top production~\cite{Li:2018tsq}, and fully differential soft functions for associated top-pair production at lepton~\cite{Liu:2024hfa} and hadron colliders~\cite{Liu:2025ldi}.

At present, however, these computations are restricted to a relatively small class of observables. For massless soft functions, by contrast, a flexible numerical framework exists that enables the computation of soft functions for essentially arbitrary global observables admitting a standard SCET factorisation theorem with hard, collinear, and soft modes. Originally developed for soft functions involving two back-to-back light-like Wilson lines~\cite{Bell:2018vaa,Bell:2018oqa,Bell:2020yzz}, the {\tt SoftSERVE} approach was subsequently extended to an arbitrary number of light-like directions in general kinematic configurations~\cite{Bell:2023yso}. Related frameworks for the computation of generic beam~\cite{Bell:2022nrj,Bell:2024epn,Bell:2024lwy} and jet functions~\cite{Bell:2021dpb,Brune:thesis} were also established thereafter.

In the present work, we extend the {\tt SoftSERVE} framework to the computation of NNLO soft functions involving \emph{time-like} Wilson lines associated with massive hard emitters. Specifically, we focus on top-quark pair production at hadron colliders, which in the partonic centre-of-mass frame features two back-to-back massive and two back-to-back massless directions. Similar to~\cite{Bell:2023yso}, we furthermore restrict our attention to SCET$_\mathrm{I}$ soft functions that satisfy the non-Abelian exponentiation theorem.

While the main result of our study is the novel framework that will significantly facilitate the computation of soft functions in the future, we also present new results for the \mbox{0-jettiness} soft function for top-quark pair production in this work. This is relevant, in particular, for applying jettiness-slicing techniques for fixed-order NNLO computations and precision resummation, as well as for the development of Monte-Carlo event generators. Specifically, the \textsc{Geneva} event generator~\cite{Alioli:2015toa} employs N-jettiness as a set of jet resolution variables for carrying out both NNLO non-local subtraction calculations~\cite{Boughezal:2015aha,Gaunt:2015pea,Alioli:2025hpa} and NNLO-accurate event generation matched to parton showers.

The \textsc{Geneva} generator was originally developed for colour-singlet production processes \cite{Alioli:2015toa,Alioli:2019qzz,Alioli:2020qrd,Alioli:2021egp, Alioli:2021qbf,Alioli:2022dkj,Alioli:2023har,Alioli:2025xcu}, and is currently being extended to colour-singlet production in association with a final-state jet. For the $Z$+jet process, in particular, N$^3$LL resummation of $1$-jettiness~\cite{Alioli:2023rxx} (neglecting coherence-violating effects \cite{Banfi:2025mra,Becher:2026kbr}) and NNLO non-local subtraction~\cite{Alioli:2025hpa} have been implemented. The generalised {\tt SoftSERVE} framework presented in this work opens the way to a further extension of \textsc{Geneva} to heavy-quark pair production. The corresponding factorisation and resummation formulae for 0-jettiness were derived in~\cite{Alioli:2021ggd}, together with the NLO soft function projected onto a specific colour basis. These results form the starting point of our analysis, and the soft function computed in this work supplies an essential ingredient required to reach NNLL$^\prime$ accuracy.

The remainder of this paper is organised as follows. Section~\ref{sec:technical} describes the technical aspects of the calculation. In particular, we discuss the choice of reference frames that simplifies the numerical implementation and present our approach to evaluating the virtual-real and double-real NNLO contributions to the soft functions. Section~\ref{sec:ren} is devoted to the renormalisation procedure within the combined SCET+Heavy Quark Effective Theory (HQET) framework. In section~\ref{sec:results}, we present numerical results for the 0-jettiness soft function and demonstrate the pole cancellation numerically, where we also derive an analytic expression for the tripole contributions in the threhold limit. Finally, our conclusions are presented in section~\ref{sec:conclusions}, and the relevant anomalous dimensions for our analysis are summarised in the appendix.

\section{Technical aspects of the calculation}
\label{sec:technical}

\subsection{Definition of the soft function}

The soft function for a process with two initial-state massless partons and a pair of final-state massive quarks is defined as a correlation function of soft Wilson lines,
\begin{align}\label{eq:softoper}
\mathbf{S} (\tau,\betat,\vartheta,\mu) = \sum_{X_s}\!\!\!\!\!\!\!\!\!\int\, \bra{0}\bar{\mathbf{T}}\big[\bm{O}^{s\, \dagger}(0)\big]\ket{X_s}\bra{X_s}\mathbf{T} \big[\bm{O}^s(0)\big]  \ket{0} \, \mathcal{M}(\tau;\{k_m\}) \, ,
\end{align}
where the soft operator is expressed as
\begin{align}
\bm{O}^s(x) = \bm{S}_{v_4}(x) \bm{S}_{v_3}(x) \bm{S}_{n_2}(x) \bm{S}_{n_1}(x) \, ,
\end{align}
and the soft Wilson lines, which extend along light-like $n_{\{1,2\}}$ and time-like $v_{\{3,4\}}$ directions, are given by
\begin{align}
\bm{S}_{n_{\{1,2\}}}(x)& = \mathbf{P} \, \mathrm{exp} \bigg[ i g_s \int^0_{-\infty}\, \de t \, n_{\{1,2\}}\cdot A^a_s(x+t \,n_{\{1,2\}})\, \mathbf{T}^a_{\{1,2\}} \bigg] \, , \nonumber\\
\bm{S}_{v_{\{3,4\}}}(x) & = \mathbf{P}\, \mathrm{exp} \bigg[ i g_s \int^\infty_{0}\, \de t \, v_{\{3,4\}}\cdot A^a_s(x+t \, v_{\{3,4\}})\, \mathbf{T}^a_{\{3,4\}} \bigg]\, .
\end{align}
The action of the colour operators $\mathbf{T}^a_i$ is defined in the colour-space formalism~\cite{Catani:1996vz}, and the symbol $\mathbf{P}$ denotes path ordering. Within this approach, the contributions from conjugate quark fields and the appropriate path ordering are handled automatically. \footnote{For an introduction to this formalism, see for example appendix~H of \cite{Becher:2014oda}.}
In \eqref{eq:softoper} the combined sum/integral sign represents the integration over the phase space of the emitted soft partons with momenta $\{k_m\}$, and the measurement function  $\mathcal{M}(\tau;\{k_m\})$  specifies what is measured on the soft radiation. In the {\tt SoftSERVE} approach~\cite{Bell:2018vaa,Bell:2018oqa,Bell:2020yzz,Bell:2023yso}, one assumes that the latter can be written in the form
\begin{align} \label{eq:measure}
\mathcal{M}(\tau;\{k_m\}) = \mathrm{exp}\bigl(- \tau \, \omega(\{k_m\})\bigr)\, ,
\end{align}
where $\tau$ is the Laplace-space conjugate variable to the observable under consideration. The measurement function is furthermore subject to certain constraints that are described in detail in \cite{Bell:2018oqa}. 

In back-to-back kinematics, which is the relevant case for hadronic $t\bar{t}$ production, the soft function has a kinematical dependence on two variables: the heavy-quark velocity in the partonic centre-of-mass frame \mbox{$\beta_t=\sqrt{1- 4 m_t^2/M^2_{t \bar{t}}}$}, where $M_{t \bar{t}}$ is the invariant mass of the $t \bar{t}$ pair, and the scattering angle $\vartheta$ between the beam axis and the direction of the top quark in the same frame. For simplicity, in what follows we suppress the dependence on $\beta_t$ and $\vartheta$, restoring it only where relevant to the discussion.

In this work, we aim at developing a framework for the numerical evaluation of soft functions with a generic measurement function \eqref{eq:measure} to NNLO in perturbation
theory. To this end, we define the perturbative coefficients of the bare soft function as
\begin{align}\label{eq:softfundef}
    \mathbf{S}_0(\tau) = \mathbf{1}+ \bigg(\frac{Z_{\alpha_s} \alpha_s}{4 \pi}\bigg)\,  (\mu^2 \bar{\tau}^2)^\epsilon \,\mathbf{S}^{(1)}(\epsilon) + \bigg(\frac{Z_{\alpha_s} \alpha_s}{4 \pi}\bigg)^2 \, (\mu^2 \bar{\tau}^2)^{2 \epsilon}  \, \mathbf{S}^{(2)}(\epsilon) +\mathcal{O}(\alpha_s^3)\,,
\end{align}
where $\epsilon = (4-d)/2$ is the dimensional regulator, $\bar{\tau}= \tau e^{\gamma_E}$, and $\alpha_s$ is the strong coupling constant in the $\overline{\mathrm{MS}}$ scheme, which is related to the bare coupling $\alpha_s^0$ via $Z_{\alpha_s} \alpha_s \mu^{2 \epsilon}= e^{-\epsilon \gamma_E} (4 \pi)^\epsilon \alpha^0_s$ where $Z_{\alpha_s}=1- \beta_0 \alpha_s/(4 \pi \epsilon)$ and $\beta_0=11/3 C_A-4/3 T_F n_f$.
The boldface quantities in \eqref{eq:softfundef} carry a non-trivial colour structure.

\subsection{Frame choices}
\label{sec:frames}

The amplitudes entering the calculation exhibit divergences associated with the limits in which massless partons become soft or collinear to one another. These limits must be made manifest in the calculation by choosing suitable parameterisations for the phase-space integrations. For generic massless soft functions, a suitable choice of variables has been introduced in~\cite{Bell:2023yso}, which we also adapt here. The form of this parameterisation was motivated primarily by the requirement that divergences due to emissions becoming collinear to either one of the emitting dipole directions must be captured by the endpoints of the integration range.
As a consequence, the parameterisation was built on a light-cone decomposition in a boosted frame, in which the emitting dipole directions are back-to-back. A particularly appealing consequence of this property is that even for dipole vectors at generic opening angles, almost no variables parameterising the transverse directions (in the boosted frame) enter the integration kernel originating from the matrix element. The full transverse-space parameterisation instead only enters through the explicit form of the measurement function, which reduces the computational complexity and, in particular, the dimension of the non-trivial integration domain as we will see below.

For the $t\bar{t}$ case considered here, there are no divergences associated with collinearity to the heavy-quark directions, and the only massless external directions are those of the beams. In principle, there are then several options for finding suitable parameterisations in this case. On the one hand, one could use a straightforward light-cone decomposition in the partonic centre-of-mass frame for all contributions of the calculation. This would make the collinear divergences associated with emissions from the beam directions manifest, but it would render the dependence on the top-quark kinematics somewhat unwieldy, as the reference vectors for the top quarks would not be aligned with any of the coordinate axes. Scalar products between the momenta of the emitted partons and the top-quark reference vectors would therefore involve all vector components and take a rather complicated form.
The advantage, however, would be that the measurement functions would usually take a particularly simple form, since the observables are often defined in this frame. Indeed, the 0-jettiness measurement function we consider in this paper is almost trivial in this parameterisation.

On the other hand, one could again establish a light-cone decomposition in a boosted frame that takes the dependence on the top-quark kinematics into account, even though the collinear divergences are absent in this case. The scalar products appearing in the master formulae would then be much simpler, since aligning the coordinate axes with the top-quark reference vectors  projects out individual components of the emission momenta. With this choice, however, the measurement functions would typically be more complicated than in the centre-of-mass frame.

The two choices represent two approaches for dealing with kinematic complexity. Choosing the centre-of-mass frame simplifies the form of the measurement function, at the expense of a more complicated matrix element. Choosing the boosted frame instead shifts the complicated functional dependence from the matrix element into the measurement function. As both options use a standard light-cone decomposition in terms of two back-to-back light-like vectors $n^\mu$ and $\bar{n}^\mu$, we can consider both options in parallel.

Starting with the massless-massless dipole in the partonic centre-of-mass frame, which already features back-to-back emitters, we parameterise the beam ($1$ and $2$) as well as the top-pair directions ($3$ and $4$), using
\begin{alignat}{2}\label{eq:para:massless}
n_1 &= (1,0,0,1)\,, &\hspace{1.5cm} n_2&=(1,0,0,-1)\, ,  \\
v_3 &= \frac{1}{\sqrt{1-\beta_t^2}}(1,\beta_t\sin\vartheta,0,\beta_t\cos\vartheta)\,, & v_4&=\frac{1}{\sqrt{1-\beta_t^2}}(1,-\beta_t\sin\vartheta,0,-\beta_t\cos\vartheta)\,,  \nonumber
\end{alignat}
which satisfy
\begin{align}
n_1 \cdot v_3 = n_2 \cdot v_4 = \frac{1-\beta_t\cos\vartheta}{\sqrt{1-\beta_t^2}}\,, \qquad\qquad
n_1 \cdot v_4 = n_2 \cdot v_3 = \frac{1+\beta_t\cos\vartheta}{\sqrt{1-\beta_t^2}}\,,
\end{align}
along with $v_3\cdot v_4 = (1+\beta_t^2)/(1-\beta_t^2)$ and $v_3^2=v_4^2=1$. The light-cone decomposition then proceeds using the vectors $n^\mu=n_1^\mu$ and $\bar{n}^\mu=n_2^\mu$, including a suitable (Euclidean) parameterisation for the transverse space,
\begin{align}\label{eq:klmomenta}
k^\mu &= k_- \frac{n^\mu}{2}  + k_+ \frac{\bar{n}^\mu}{2}+ k_x \,e_x^\mu + k_y \,e_y^\mu + \ldots 
\end{align}
where $e_x^\mu=(0,\vec{e}_x)$ and $e_y^\mu=(0,\vec{e}_y)$ are the standard unit vectors spanning the physical dimensions of the transverse space,  while any unphysical dimensions arising due to the use of dimensional regularisation are hiding in the dots.

For the massive-massive dipole, on the other hand, we can either use the light-cone decomposition in the partonic centre-of-mass frame, as mentioned above, or we rotate to a coordinate system in which the top pair is aligned with the $z$-axis. In the rotated frame, the reference directions become
\begin{alignat}{2}\label{eq:para:massive}
n'_1 &= (1,-\sin\vartheta,0,\cos\vartheta)\,,&\hspace{1.5cm} n'_2&=(1,\sin\vartheta,0,-\cos\vartheta) \, ,\nonumber\\
v'_3 &= \frac{1}{\sqrt{1-\beta_t^2}}(1,0,0,\beta_t)\,, & v'_4&=\frac{1}{\sqrt{1-\beta_t^2}}(1,0,0,-\beta_t)\,, 
\end{alignat}
and we use the same light-cone decomposition \eqref{eq:klmomenta} in terms of $n^\mu=(1,0,0,1)$ and $\bar{n}^\mu=(1,0,0,-1)$, whose spatial components point into the directions of the top quarks in this case.

Finally, for the mixed massless-massive dipole, we can also use the light-cone decomposition in the centre-of-mass frame, or we boost to a frame, in which both the massless and the massive emitter are aligned along the $z$-axis. For the (13) dipole, for instance, this is realised by
\begin{align}\label{eq:para:mixed}
n_1'' &= (1,0,0,1)\, ,\nonumber\\ 
n_2''&=\bigg(\frac{\beta_t^2s^2}{(1-\beta_t c)^2}+1,\frac{2\beta_t s}{1-\beta_t c},0,\frac{\beta_t^2s^2}{(1-\beta_t c)^2}-1\bigg)\, , \nonumber \\
v_3'' &= \frac{1}{\sqrt{1-\beta_t^2}} \bigg(1- \frac{\beta_t^2 s^2}{2(1-\beta_t c)},0,0,\beta_t c-\frac{\beta_t^2 s^2}{2(1-\beta_t c)}\bigg)\, , \nonumber\\
v_4''&= \frac{1}{\sqrt{1-\beta_t^2}} \bigg(
1+ \frac{\beta_t^2 s^2(3-\beta_t c)}{2(1-\beta_t c)^2},\frac{2\beta_t s}{1-\beta_t c}, 0,-\beta_t c+ \frac{\beta_t^2 s^2(3-\beta_t c)}{2(1-\beta_t c)^2}\bigg) \,, 
\end{align}
where we introduced the shorthand notation $s=\sin\vartheta$ and $c=\cos\vartheta$ to simplify the formulae. In this frame, we then again introduce standard light-cone coordinates using $n^\mu=(1,0,0,1)$ and $\bar{n}^\mu=(1,0,0,-1)$ along the axis of the two emitters.

The advantage of this procedure is that the kinematic scalar products entering the matrix elements take a particularly simple form in all cases. Denoting the momentum of the emitted parton by $k^\mu$, the (12) dipole depends e.g.~on the scalar products $k\cdot n_1$ and $k\cdot n_2$, which can be expressed exclusively via the $k_+$ and $k_-$ variables in \eqref{eq:klmomenta}. As can easily be verified, this is also true for the relevant scalar products for the (34) dipole using \eqref{eq:para:massive} and the (13) dipole using \eqref{eq:para:mixed}. The transverse-space variables thus do not appear in the matrix elements, merely in the measurement function.

Before continuing with the technical description of the computational method, we introduce the $0$-jettiness event shape in heavy-quark pair production~\cite{Alioli:2021ggd} that we will address in detail in section~\ref{sec:results}. In these conventions, the $0$-jettiness is defined in the partonic centre-of-mass frame as
\begin{align}
    \Tau_0 = \sum_m \min\big\{ n_1\cdot k_m, \, n_2\cdot k_m \big\} \, ,
\end{align}
where $n_1^\mu$ and $n_2^\mu$ are the light-like vectors along the two beam directions defined in~\eqref{eq:para:massless} and $k_m^\mu$ denote the momenta of the soft emissions.

\subsection{NLO contribution}
 
Within the \texttt{SoftSERVE} approach, one starts from a general parameterisation of the measure\-ment function \eqref{eq:measure} that covers a large class of observables. Specifically, for a single emission with momentum $k^\mu$, one writes
\begin{equation}\label{eq:1emMeas}
    \!\mathcal{M}_1(\tau;k) = \exp\left\lbrace-\tau k_T \left(y_k^{n/2}f_A(y_k,t_k)\,\theta(1-y_k)+y_k^{-n/2}f_B(y_k^{-1},t_k)\,\theta(y_k-1)\right)\right\rbrace ,
\end{equation}
where we expressed the light-cone components in \eqref{eq:klmomenta} in the chosen frame as
\begin{align}
    k_+ &=k_T\, \sqrt{y_k}\, ,\hspace{2cm} 
    k_-=\frac{k_T}{\sqrt{y_k}}\, , \nonumber
    \\
    k_x &=k_T \,\cos \theta_{k1}\, , \qquad\qquad
    k_y=k_T \,\sin \theta_{k1} \cos \theta_{k2}\, , \qquad\qquad
    \ldots 
\end{align}
and the dots refer to further transverse directions that are parameterised by $(d-2)$ dimensional spherical coordinates in the usual way. As is clear from the discussion in  section~\ref{sec:frames}, we can choose the coordinate system such that the physical reference vectors have vanishing $y$-component, and therefore the angle $\theta_{k2}$ -- as well as all angles that parameterise the unphysical $(d-4)$ directions -- do not enter the measurement function \eqref{eq:1emMeas}. The angular dependence of the measurement function is thus captured by a single variable, for which we choose $t_k= (1-\cos\theta_{k1})/2$. We note, however, that the situation would be different for associated $t\bar{t}$ production with non-back-to-back top quarks.

The variable $y_k$, on the other hand, parameterises collinearity to either of the two (massless or massive) emitters, and in \eqref{eq:1emMeas} we have split its natural range of $y_k\in[0,\infty)$ at $y_k=1$ to remap the upper range back to $y_k\in[0,1]$ in a subsequent step. As a result, collinearity to either of the two emitters is  encoded in the limit  $y_k\to0$ in both branches $A$ and $B$. In addition, we have introduced a parameter $n$, which in the massless case is related to the relative power counting of the soft and collinear modes (see~\cite{Bell:2018oqa} for details). On the level of the calculation, the value of $n$ must be properly set to extract the correct coefficient of the collinear singularities by adjusting it such that the functions $f_{A/B}$ stay finite in the limit  $y_k\to0$. As the collinear divergences are absent for massive emitters, there is no need to factor out an appropriate power of $y_k$ in this case, but we prefer to do so with the same value of $n$ for these contributions as well. As we will see explicitly below, the resulting functions $f_{A/B}$ are then not necessarily finite in the limit  $y_k\to0$, which is however also not required due to the absence of collinear divergences for massive partons. We also note that we focus on SCET$_\mathrm{I}$ observables with $n\neq0$ in this work.

At NLO the one-loop virtual corrections yield scaleless integrals in our setup, and one is left with single-emission corrections. More precisely, the NLO coefficient of the bare soft function in \eqref{eq:softfundef} receives dipole and monopole contributions,
\begin{equation}
  \mathbf{S}^{(1)}(\epsilon) = \sum_{\alpha,\beta} \,\mathbf{T}_\alpha\cdot\mathbf{T}_\beta \ \tilde{S}^{(1)}_{\alpha\beta}(\epsilon) \, , 
\end{equation}
with the sum over the greek indices $\alpha$, $\beta$ running over both massless and massive legs, and
\begin{equation}
    \tilde{S}^{(1)}_{\alpha\beta}(\epsilon) = -\frac{\left(4\pi e^{\gamma_E} \tau^2\right)^{-\epsilon}}{(2\pi)^{d-1}}\int \df^d k \ \delta(k^2)\theta(k^0)\,\mathcal{M}_1(\tau;k)\,\abs{\tilde{\mathcal{A}}_{\alpha\beta}(k)}^2 \, .  
\end{equation}
Using colour conservation, it is possible to express the square of the NLO soft matrix element solely in terms of dipole contributions,
\begin{align}\label{eq:NLOsoftamp}
\sum_{\alpha,\beta}   \,\abs{\tilde{\mathcal{A}}_{\alpha\beta}(k)}^2 \;   \,\mathbf{T}_{\alpha} \cdot \mathbf{T}_{\beta}   &= \sum_{\alpha \neq \beta} \bigg(\abs{\tilde{\mathcal{A}}_{\alpha\beta}(k)}^2 - \frac{\abs{\tilde{\mathcal{A}}_{\alpha\alpha}(k)}^2}{2} - \frac{\abs{\tilde{\mathcal{A}}_{\beta\beta}(k)}^2}{2} \bigg) \, \mathbf{T}_\alpha \cdot \mathbf{T}_\beta  \, \nonumber \\
     &= \sum_{\alpha < \beta} \big(2\, \abs{\tilde{\mathcal{A}}_{\alpha\beta}(k)}^2 - \abs{\tilde{\mathcal{A}}_{\alpha\alpha}(k)}^2 - \abs{\tilde{\mathcal{A}}_{\beta\beta}(k)}^2 \big) \; \mathbf{T}_\alpha \cdot \mathbf{T}_\beta \nonumber\\
     & \equiv \sum_{\alpha < \beta} \,\abs{\mathcal{A}_{\alpha \beta}(k)}^2 \; \mathbf{T}_\alpha \cdot \mathbf{T}_\beta \, ,  
\end{align}
where 
\begin{align}\label{eq:NLO:matrixelement}
    \abs{\tilde{\mathcal{A}}_{\alpha\beta}(k)}^2 = 16\pi^2\frac{v_\alpha \cdot v_\beta}{(v_\alpha\cdot k) (v_\beta \cdot k)}\equiv 16\pi^2 e_{\alpha\beta} \, ,  
\end{align}
and $v^\mu_\alpha, v^\mu_\beta$ represent the light-like or time-like directions from above.  It is sometimes convenient to introduce a slightly different notation by separating massless and massive indices using lowercase $i,j$ and uppercase $I,J$ latin letters respectively. Hence we can rewrite the last line in \eqref{eq:NLOsoftamp} as
\begin{align}
& \sum_{\alpha< \beta}  \, \abs{\mathcal{A}_{\alpha\beta}(k)}^2 \; \mathbf{T}_{\alpha} \cdot \mathbf{T}_{\beta}  \nonumber \\
& \qquad =\sum_{i < j} \, \abs{\mathcal{A}_{ij}(k)}^2\;  \mathbf{T}_i \cdot \mathbf{T}_j + \sum_{i,J} \,\abs{\mathcal{A}_{iJ}(k)}^2  \;\mathbf{T}_i \cdot \mathbf{T}_J  + \sum_{I < J} \, \abs{\mathcal{A}_{IJ}(k)}^2 \;\mathbf{T}_I \cdot \mathbf{T}_J  \, . 
\end{align}
We will occasionally switch between generic indices $\alpha,\beta$ and mass-specific indices $i,J$ in the following.

Making the dipole structure explicit, one then starts from the master formula 
\begin{equation}\label{eq:nlodipole}
    \mathbf{S}^{(1)}(\epsilon) = \sum_{\alpha<\beta} \,\mathbf{T}_\alpha\cdot\mathbf{T}_\beta \ S^{(1)}_{\alpha\beta}(\epsilon) \, , 
\end{equation}
where the monopole contributions have been absorbed into the dipole coefficients $S^{(1)}_{\alpha\beta}(\epsilon)$. We then introduce the $k_T$-stripped matrix element
\begin{equation}
    \mathcal{J}_{\alpha\beta}(y_k,t_k)=k_T^2 \, \frac{\abs{\mathcal{A}_{\alpha\beta}(k)}^2}{64\pi^2}\, ,
\end{equation}
and perform the integration over $k_T$ as well as all angles that do not appear in the measure\-ment function to arrive at the following master formula for the calculation of NLO bare soft functions,
\begin{align}\label{eq:NLO:master}
    S_{\alpha\beta}^{(1)}(\epsilon) &= -\frac{8e^{-\gamma_E\epsilon}}{\sqrt{\pi}}\frac{\Gamma(-2\epsilon)}{\Gamma(1/2-\epsilon)}\,\int_0^1\df t_k \, \big(4t_k\bar t_k\big)^{-1/2-\epsilon} \nonumber\\
    &\qquad \times \int_0^1 \frac{\df y_k}{y_k^{1-n\epsilon}}\;\Big[ f_A(y_k,t_k)^{2\epsilon}\,\mathcal{J}_{\alpha\beta}(y_k,t_k)+f_B(y_k,t_k)^{2\epsilon}\,\mathcal{J}_{\alpha\beta}(y_k^{-1},t_k)\Big]\, , 
\end{align}
where $\bar t_k=1-t_k$. Since we use light-cone coordinates in suitable frames for all emitting dipoles as explained above, we simply need to account for the different forms the matrix elements $\mathcal{J}_{\alpha\beta}$ and the measurement functions $f_{A/B}$ attain in the different frames.

For the massless-massless dipole, this exercise is trivial and the calculation matches that of massless soft functions in~\cite{Bell:2018oqa} with $\mathcal{J}_{12}(y_k,t_k) = 1$. The $y_k$-integral in \eqref{eq:NLO:master} encapsulates a collinear divergence in this case  in the limit $y_k\to 0$ in both the $A$ and $B$ regions, when the emitted gluon aligns with either one of the dipole directions.

In the massless-massive case, e.g.~the $(13)$ dipole, we find in the parameterisation \eqref{eq:para:mixed}
\begin{align} \label{eq:me:nlo:mixed}
    \mathcal{J}_{13}(y_k,t_k) &= \frac{(1-\beta_t \cos \vartheta)^4}{\big[(1-\beta_t\cos\vartheta)^2+y_k\,(1-\beta_t^2)\big]^2}\,, 
\end{align}
which is again finite in the limit $y_k\to 0$, generating a collinear divergence in region $A$ when the gluon becomes collinear to the massless dipole direction. In region $B$, on the other hand, the matrix element $\mathcal{J}_{13}(y_k^{-1},t_k)$ suppresses the divergence in the limit $y_k\to 0$ as there is no divergence associated with collinearity to massive dipole directions.

Finally, for the massive-massive dipole we have in the parameterisation \eqref{eq:para:massive}
\begin{align} \label{eq:me:nlo:massive}
    \mathcal{J}_{34}(y_k,t_k) &= \frac{16\beta_t^2y_k^2}{\big[(1+y_k)^2-(1-y_k)^2 \beta_t^2\big]^2}\,, 
\end{align}
where in both regions $A$ and $B$ the collinear divergence in the limit $y_k\to0$ is suppressed by the matrix element, as the collinear divergences are absent in this case.

Besides the obvious symmetry $\abs{\mathcal{A}_{\alpha\beta}}^2=\abs{\mathcal{A}_{\beta\alpha}}^2$ due to \eqref{eq:NLO:matrixelement}, there are also symmetries originating from the back-to-back kinematics, which after integration lead to the following relations:
\begin{equation}\label{eq:NLO:symmetries}
    S_{13}^{(1)}(\epsilon;\vartheta) = S_{24}^{(1)}(\epsilon;\vartheta)= S_{14}^{(1)}(\epsilon;\pi-\vartheta)= S_{23}^{(1)}(\epsilon;\pi-\vartheta)\,. 
\end{equation}
It is therefore sufficient to calculate one mixed dipole explicitly, along with the massive-massive dipole, since the purely massless one can directly be computed using the public {\tt SoftSERVE} distribution. \footnote{The {\tt SoftSERVE} package is publicly available at \url{https://softserve.hepforge.org/}.}

While this framework applies to general observables that can be written in the form \eqref{eq:1emMeas}, we will present numerical results for 0-jettiness in section \ref{sec:results} as an explicit proof of concept. For this observable, one has $n=1$ and the measurement functions become in the parameterisations from section \ref{sec:frames},
\begin{align}
    f_A^{(12)}(y_k,t_k)&=f_B^{(12)}(y_k,t_k)=1\, ,\nonumber\\
    f_A^{(13)}(y_k,t_k)&=\min\Bigl(1,\frac{1}{y_k}-\frac{2\beta_t(1-2t_k)\sin\vartheta}{\sqrt{y_k}(1-\beta_t \cos \vartheta)}+\frac{\beta_t^2\sin^2\vartheta}{(1-\beta_t\cos\vartheta)^2}\Bigr)\, ,\nonumber\\
    f_B^{(13)}(y_k,t_k)&=\min\Bigl(\frac{1}{y_k},1-\frac{2\beta_t(1-2t_k)\sin\vartheta}{\sqrt{y_k}(1-\beta_t \cos \vartheta)}+\frac{\beta_t^2\sin^2\vartheta}{y_k(1-\beta_t\cos\vartheta)^2}\Bigr)\, , 
    \label{eq:0-jettiness:M1}\\
    f_A^{(34)}(y_k,t_k)&= f_B^{(34)}(y_k,t_k)     \nonumber \\
    &=\min\Bigl(\frac{\cos^2\frac{\vartheta}{2}}{y_k}+\sin^2\frac{\vartheta}{2}+\frac{(1-2t_k)\sin\vartheta}{\sqrt{y_k}},\frac{\sin^2\frac{\vartheta}{2}}{y_k}+\cos^2\frac{\vartheta}{2}-\frac{(1-2t_k)\sin\vartheta}{\sqrt{y_k}}\Bigr)\, . \nonumber
\end{align}
These expressions allow us to verify the scaling in the limit $y_k\to0$ explicitly. Specifically, we observe that the functions $f_{A/B}^{(12)}$ and $f_{A}^{(13)}$ are finite in this limit, as required for the extraction of the associated collinear singularities. The functions $f_B^{(13)}$ and $f_{A/B}^{(34)}$, on the other hand, are not finite, which does not pose a problem since the collinear divergences are absent in this case.

\subsection{NNLO contribution}

At NNLO, the soft function receives double-virtual, real-virtual, and double-real corrections. The double-virtual corrections again yield scaleless integrals and vanish in dimensional regularisation; this vanishing is precisely what converts the infrared poles into ultra\-violet poles, which are subsequently removed by renormalisation. The real-virtual and double-real corrections, by contrast, are non-zero, and their computation is described in the subsections below.
For this reason, we decompose the second-order coefficient of the soft function introduced in \eqref{eq:softfundef} into its non-zero contributions,
\begin{align}\label{eq:genericNNLO}
    \mathbf{S}^{(2)}(\epsilon) = \mathbf{S}^{(2,\text{RV})}(\epsilon) + \mathbf{S}^{(2,gg)}(\epsilon) +\mathbf{S}^{(2,q\bar{q})}(\epsilon)\, , 
\end{align}
where $\mathbf{S}^{(2,\text{RV})}(\epsilon)$, $\mathbf{S}^{(2,\text{gg})}(\epsilon)$, and $\mathbf{S}^{(2,q\bar{q})}(\epsilon)$ denote the real-virtual, double-real gluon, and double-real fermion contributions, respectively.

As far as the phase-space integration is concerned, the real-virtual corrections are of a complexity comparable to the NLO calculation, since only a single real emission has to be integrated over. This holds provided that the one-loop amplitude is known analytically, as it is in our case. The double-real corrections, on the other hand, constitute the most complicated part of the computation, and we again follow the {\tt SoftSERVE} approach to evaluate the respective phase-space integrals.

In terms of colour structures, the real-virtual term receives contributions from dipole and tripole colour correlations, whereas the double-real term has a structure consisting of dipole and quadrupole correlations. For the class of observables considered in this work, namely those that obey non-Abelian exponentiation, the quadrupole colour correlations originate entirely from the exponentiation of the NLO corrections. \footnote{Naively one might expect there to also be tripole contributions from exponentiation. However, these contributions vanish due to their colour structures.} We therefore do not need to calculate these, and they will e.g.~not appear in the renormalisation section below, which we implement using the logarithm of the soft function, assuming that non-Abelian exponentiation holds.

\subsubsection{Real-virtual interference}

The real-virtual contribution to the NNLO soft function involves the same one-emission measurement function in \eqref{eq:1emMeas} as the NLO calculation, and therefore only requires a parameterisation of the one-particle phase space. Contrary to NLO, however, it has both a dipole contribution -- after absorbing the monopoles similar to \eqref{eq:NLOsoftamp} -- as well as a tripole contribution,
\begin{align}\label{eq:rvcoloursum}
    \mathbf{S}^{(2,\text{RV})}(\epsilon)&= C_A\sum_{\alpha < \beta}\,\mathbf{T}_\alpha\cdot\mathbf{T}_\beta \ S^{(2,\mathrm{Re})}_{\alpha\beta}(\epsilon)+\sum_{\alpha\neq\beta\neq\gamma}f^{abc}\mathbf{T}^a_\alpha\mathbf{T}^b_\beta\mathbf{T}^c_\gamma \ S^{(2,\mathrm{Im})}_{\alpha\beta\gamma}(\epsilon)\, , 
\end{align}
where
\begin{align}
S^{(2,\mathrm{Re})}_{\alpha\beta}(\epsilon) &= \frac{16\pi^2(4\pi e^{3\gamma_E} \tau^4)^{-\epsilon}}{(2\pi)^{d-1}}\int \, d^dk \, \delta(k^2) \theta(k^0) \, \mathcal{M}_1(\tau;k)\, \nonumber\\
& \qquad \qquad \qquad \qquad \qquad \times \left[2 \, \left(\frac{e_{\alpha\beta}}{2}\right)^\epsilon(2e_{\alpha\beta}-e_{\alpha\alpha}-e_{\beta\beta})R_{\alpha\beta}(\epsilon)\right], 
\label{eq:RV}\\
S^{(2,\mathrm{Im})}_{\alpha\beta\gamma}(\epsilon) &= \frac{16\pi^2(4\pi e^{3\gamma_E} \tau^4)^{-\epsilon}}{(2\pi)^{d-1}}\int \, d^dk \,  \delta(k^2) \theta(k^0) \, \mathcal{M}_1(\tau;k)\, \left[-4\pi \, e_{\alpha\gamma}\left(\frac{e_{\alpha\beta}}{2}\right)^\epsilon I_{\alpha\beta}(\epsilon)\right]\,,
\nonumber
\end{align}
which depend on the eikonal factors $e_{\alpha\beta}$ defined in \eqref{eq:NLO:matrixelement}, as well as the real part  $R_{\alpha \beta}(\epsilon)$ and imaginary part $I_{\alpha \beta}(\epsilon)$ of the one-loop soft matrix element respectively. More specifically, we use the results for the massive one-loop soft current\footnote{The tripole contributions also involve the massless one-loop soft current that was first computed in~\cite{Catani:2000pi}.} from \cite{Czakon:2018iev} -- which uses slightly different conventions than \cite{Bierenbaum:2011gg} -- after checking them both numerically and analytically. Compared to the NLO calculation, the real-virtual contribution thus adds additional functions to the integrand, which do not affect the divergence structure and can therefore be dealt with straightforwardly.

For the back-to-back kinematics considered here, exchanging either the two beam or the two heavy-quark directions amounts to replacing $\vartheta\to\pi-\vartheta$. For the dipoles, we have already used this property at NLO in~\eqref{eq:NLO:symmetries}. The symmetry applies equally at higher orders for both dipoles and tripoles, and for the latter it implies e.g.~that
\begin{equation}
    S^{(2,\mathrm{Im})}_{123}(\epsilon;\vartheta) =S^{(2,\mathrm{Im})}_{214}(\epsilon;\vartheta)=S^{(2,\mathrm{Im})}_{124}(\epsilon;\pi-\vartheta)= S^{(2,\mathrm{Im})}_{213}(\epsilon;\pi-\vartheta) \,.
\end{equation}
In total, there are hence only three independent dipoles as explained above, and six independent tripoles, e.g.~(123), (132), (134), (312), (314) and (341). Furthermore, due to colour conservation, there is only one independent combination of three colour charge operators $f^{abc}\mathbf{T}_\alpha^a\mathbf{T}_\beta^b\mathbf{T}_\gamma^c$, e.g. $(\alpha\beta\gamma)=(123)$, so that we can write the total tripole contribution to the bare NNLO soft function as:
\begin{align}\label{eq:tripsum}
    & \sum_{\alpha\neq\beta\neq\gamma}f^{abc}\mathbf{T}^a_\alpha\mathbf{T}^b_\beta\mathbf{T}^c_\gamma \ S^{(2,\mathrm{Im})}_{\alpha\beta\gamma}(\epsilon;\vartheta) \\
    & \quad =2\, f^{abc}\mathbf{T}_1^a\mathbf{T}_2^b\mathbf{T}_3^c \; \bigg[
    S^{(2,\mathrm{Im})}_{123}(\epsilon;\vartheta)-S^{(2,\mathrm{Im})}_{123}(\epsilon;\pi-\vartheta)+S^{(2,\mathrm{Im})}_{341}(\epsilon;\vartheta)-S^{(2,\mathrm{Im})}_{341}(\epsilon;\pi-\vartheta)\nonumber\\
     & \hspace{35mm}-S^{(2,\mathrm{Im})}_{132}(\epsilon;\vartheta)+S^{(2,\mathrm{Im})}_{132}(\epsilon;\pi-\vartheta)+S^{(2,\mathrm{Im})}_{312}(\epsilon;\vartheta)-S^{(2,\mathrm{Im})}_{312}(\epsilon;\pi-\vartheta)\nonumber\\
     &\hspace{35mm}-\, S^{(2,\mathrm{Im})}_{314}(\epsilon;\vartheta)+S^{(2,\mathrm{Im})}_{314}(\epsilon;\pi-\vartheta)+S^{(2,\mathrm{Im})}_{134}(\epsilon;\vartheta)-S^{(2,\mathrm{Im})}_{134}(\epsilon;\pi-\vartheta)\bigg]. \nonumber
\end{align}
We close this subsection with a remark. The one-loop soft current from \cite{Bierenbaum:2011gg,Czakon:2018iev} is provided as an expansion up to $\mathcal{O}(\epsilon^2)$, which together with the soft and collinear divergences of the phase-space integrations yields a finite contribution to the NNLO soft function. As the collinear divergences are absent for the massive-massive dipole, the real part $R_{34}(\epsilon)$ has only been computed up to $\mathcal{O}(\epsilon)$ in \cite{Bierenbaum:2011gg,Czakon:2018iev}. The situation is, however, subtle for the corresponding tripoles, which are sensitive to the imaginary part $I_{34}(\epsilon)$. Specifically, when the index $\gamma$ in \eqref{eq:RV} refers to a massless direction, the phase-space integrals yield soft \emph{and} collinear divergences, requiring the $\mathcal{O}(\epsilon^2)$ terms of $I_{34}(\epsilon)$. This piece has not been provided in the literature.

Closer inspection reveals, however, that this contribution cancels in the sum over all tripoles and therefore results derived on the basis of the one-loop current in \cite{Bierenbaum:2011gg,Czakon:2018iev} are complete at NNLO. The reason is that the $\mathcal{O}(\epsilon^2)$ coefficient of the current requires both soft and collinear divergences from the phase-space integrals, and in this limit the matrix element loses its dependence on the kinematic configuration of the tripole. In the tripole sum \eqref{eq:tripsum}, such contributions drop out, as is e.g.~visible in the combination $S^{(2,\mathrm{Im})}_{341}(\vartheta)-S^{(2,\mathrm{Im})}_{341}(\pi-\vartheta)$. This pattern persists also for other processes such as associated $t\bar{t}$ production. To the best of our knowledge, this particularity has not been emphasised in the literature before.

\subsubsection{Double real emissions}

The double real-emission corrections in~\eqref{eq:genericNNLO} receive contributions from two channels, the emission of gluon-gluon ($gg$) and quark-antiquark ($q\bar{q}$) pairs.
The gluonic contribution involves both dipole and quadrupole colour structures (after absorbing the monopoles), whereas the quark-antiquark channel involves only colour dipoles:
\begin{align}\label{eq:colorsep}
\mathbf{S}^{(2,gg)}(\epsilon)&=C_A \, \sum_{\alpha<\beta}\, \mathbf{T}_\alpha\cdot\mathbf{T}_\beta \ S^{(2,gg)}_{\alpha\beta}(\epsilon)\nonumber+\frac{1}{2} \bigl(\mathbf{S}^{(1)}(\epsilon)\bigr)^2\nonumber\, ,\\
\mathbf{S}^{(2,q\bar{q})}(\epsilon)&=T_F \, n_f \, \sum_{\alpha<\beta}\, \mathbf{T}_\alpha\cdot\mathbf{T}_\beta \ S^{(2,q\bar{q})}_{\alpha\beta}(\epsilon) \,. 
\end{align}
Note that the non-Abelian exponentiation contribution. i.e.~the second term on the first line of \eqref{eq:colorsep}, also contains dipole terms, which are however uncorrelated emissions from exponentiating Abelian contributions. The first term in the first line of~\eqref{eq:colorsep} therefore describes the pure non-Abelian, correlated emission of two gluons. This is the only contribution we will calculate for gluon-pair emission, with the remainder recovered by squaring the NLO correction. 

We then follow the strategy laid out in the NLO section, specifying first the appropriate frame for the dipole under consideration. The calculation, in fact, becomes similar to the massless one covered in the original \texttt{SoftSERVE} papers~\cite{Bell:2018oqa,Bell:2023yso}. In the parameterisations from section \ref{sec:frames}, the matrix elements take a relatively simple form and the non-trivial angles parameterising the transverse space only show up in the measurement function, as we will see explicitly below. The phase-space divergences are finally factorised into monomials that can easily be expanded in terms of distributions, and the only added difficulty due to the massive emitters arises in form of a more involved  functional dependence of the integrands that are  passed to the numerical integrator. 

To make the procedure more concrete, we  apply the strategy to the quark-antiquark channel here. We thus start with
\begin{align} \label{eq:DR-qq-S}
 S_{\alpha\beta}^{(2,q\bar{q})}(\epsilon) &=  -\frac{(4 \pi e^{\gamma_E} \tau^2)^{-2\epsilon}}{(2\pi)^{2d-2}}
 \int {d}^d k \ \delta(k^2)\theta(k^0) \!
  \int {d}^d l \  \delta(l^2)\theta(l^0)  \,\mathcal{M}_{2}(\tau;k,l) \, \abs{\mathcal{A}^{q\bar{q}}_{\alpha\beta}(k,l)}^2,
\end{align}
where $k^\mu$ and $l^\mu$ are the momenta of the emitted partons, and the two-particle measurement function now takes the form \cite{Bell:2023yso}
\begin{align} \label{eq:2emMeas}
\mathcal{M}_2(\tau; k,l) &= \exp\bigg\{-\tau \,p_{T}
\bigg( 
y^{n/2}\, F_A\big(a,b,y,\{\theta\}\big) \,\Theta_A
+ y^{n/2} \, F_B\big(a,b^{-1},y,\{\theta\}\big) \,\Theta_B\\
&\qquad\qquad
+ 
y^{-n/2}\, F_C\big(a,b,y^{-1},\{\theta\}\big) \,\Theta_C
+ y^{-n/2}\, F_D\big(a,b^{-1},y^{-1},\{\theta\}\big) \,\Theta_D
\bigg)\bigg\}, \nonumber
\end{align}
where the phase space is split into four separate  contributions with $\Theta_A=\theta(1-b)\,\theta(1-y)$, $\Theta_B=\theta(b-1)\,\theta(1-y)$, $\Theta_C=\theta(1-b)\,\theta(y-1)$, and $\Theta_D=\theta(b-1)\,\theta(y-1)$. The variables appearing in the measurement function are  defined as
\begin{align}
p_T =  \sqrt{(k_{+} + l_{+})(k_{-} + l_{-})} \,, \quad\;\;\;
y &= \frac{k_{+} + l_{+}}{k_{-} + l_{-}} \,, 
\quad\;\;\;
a = \sqrt{\frac{k_{-}l_{+}}{k_{+}l_{-}}} \,,
\quad\;\;\;
b =\sqrt{\frac{k_{+}k_{-}}{l_{+}l_{-}}} \,, 
\end{align}
together with some angular variables collectively denoted by $\{\theta\}$ in \eqref{eq:2emMeas}. Note also that we factor the same power $y^{n/2}$ as at NLO to make the remnant function $F_A$ finite in the collinear limit $y\to 0$ (for massless emitters). The same holds for the other three regions.

The calculation then follows closely the lines of the massless soft-function calculation described in~\cite{Bell:2018oqa, Bell:2023yso}. Due to the back-to-back kinematics of the $t\bar t$ pair, we saw at NLO that the measurement function depends only on one non-trivial angle in the transverse plane, which gets promoted to three angles at NNLO. The situation is in this sense closer to the dijet case studied in~\cite{Bell:2018oqa} than to the general case of~\cite{Bell:2023yso}, which involves more angles that would be relevant e.g.~for associated $t\bar t$ production with non-back-to-back kinematics. Furthermore, the master formulae in~\cite{Bell:2023yso} are expressed in terms of certain kernel functions that effectively encode the corresponding matrix elements. In the massless case, these  kernels turned out to be identical in the four subregions of the integration domain $A$ through $D$ as a consequence of a symmetry: for two massless emitters, the matrix element is invariant under the exchange of the  reference vectors $n_i \leftrightarrow n_j$ as well as the parton momenta $k\leftrightarrow l$. Whereas the combined symmetry remains for the massive-massive dipole, only the $k\leftrightarrow l$ symmetry holds for the mixed massless-massive ones, and therefore one obtains different kernels for regions $A,B$ and $C,D$ in this case.

Without going into further details, we present the master formula for the quark-antiquark emission here, which reads
\begin{align}\label{eq:qq-master}
 S_{\alpha\beta}^{(2,q\bar{q})}(\epsilon) &=   
\frac{4^{-1-3\epsilon}e^{-2 \gamma_E\epsilon}\,\Gamma(1/2-2\epsilon)}{\pi^{5/2}}
 \int_0^{1} {d}y\,\int_{0}^1\,{d}a \,\int_{0}^1\, {d}b  \;a^{-2\epsilon}\, b^{-2 \epsilon}\,(a+b)^{2\epsilon}\,(1+a b)^{2\epsilon} \,
 \nn \\ & \quad 
 \int_{0}^1 {d}t_{l}\, (4 t_{l}\bar{t}_{l})^{-1/2-\epsilon}\int_{0}^1 \,{d} t'_{5} \,  (t'_{5}(2-t'_{5}))^{-1-\epsilon} 
 \int_{0}^1 \,{d} t_{kl} \,(4 t_{kl}\bar{t}_{kl})^{-1/2-\epsilon}
 \nn \\[0.2em]
 &\quad y^{-1+2n\epsilon}\, \Bigl[\mathcal{F}_1^{\alpha\beta}(a,b,y,t_{kl},t_{l},t'_{5})+\mathcal{F}_2^{\alpha\beta}(a,b,y,t_{kl},t_{l},t'_{5})\Bigr]\,,
\end{align}
with 
\begin{align}
    \mathcal{F}_1^{\alpha\beta}(a,b,y,t_{kl},t_{l},t'_{5}) &= k_{AB}^{\alpha\beta}(a,b,y,t_{kl})\Bigl(F_A(a,b,y,t_{kl},t_{l},t'_{5})^{4\epsilon}+F_B(a,b,y,t_{kl},t_{l},t'_{5})^{4\epsilon}\Bigr) \nonumber\\
    &+ k_{CD}^{\alpha\beta}(a,b,y,t_{kl})\Bigl(F_C(a,b,y,t_{kl},t_{l},t'_{5})^{4\epsilon}+F_D(a,b,y,t_{kl},t_{l},t'_{5})^{4\epsilon}\Bigr), \nonumber\\
    \mathcal{F}_2^{\alpha\beta}(a,b,y,t_{kl},t_{l},t'_{5}) &= \mathcal{F}_1^{\alpha\beta}(a,b,y,t_{kl},t_{l},2-t'_{5}) \,,
\end{align}
where the angular variables $\{t_{kl},t_l, t'_5\}$ refer to the conventions in~\cite{Bell:2018oqa}. There one can also find an explanation for the appearance of the two branches $\mathcal{F}_{1,2}^{\alpha\beta}$: they originate from a remapping of an angular variable in order to make a (spurious) singularity manifest.  It is furthermore important to note that the measurement functions $F_{A,B,C,D}$ introduced in~\eqref{eq:2emMeas} are the only functions that can depend on the entire set of integration variables in the chosen parameterisations. The kernel functions, by contrast, only depend on one angular variable, $t_{kl}$, due to our frame choice.  This simplifies the calculation, as the expansion of the measurement function is of the form $F_X^{4\epsilon}=1 + \mathcal{O}(\epsilon)$, i.e.~the dependence on the additional variables is suppressed by one order in $\epsilon$. This can lead to improved numerics, as the integrands can sometimes become independent of some of the integration variables.

The master formula for the gluon channel takes the same form with a different set of kernel functions $k'$. These kernels can be extracted from the matrix elements provided in~\cite{Catani:1999ss,Czakon:2011ve}. While the expressions are too long to be shown explicitly for the gluon channel, we provide those of the fermionic contribution for illustration here. 

The kernel for the massless-massless dipole was already given in \cite{Bell:2018oqa,Bell:2023yso}, which we repeat here for convenience,
\begin{align}\label{eq:kAB12}
k_{AB}^{12}(a,b,y,t_{kl})&=
\frac{128a}{(a+b)^2(1+a b)^2} \;\bigg\{
\frac{b (1-a^2)^2}{[(1-a)^2+4a t_{kl}]^2}
- \frac{(a+b)(1+ab)}{(1-a)^2+4a t_{kl}} \bigg\}\,,
\end{align}
with the same expression in regions $C$, $D$ because of the exchange symmetry mentioned above.

The same symmetry holds for the massive-massive kernel, which in the parameterisation \eqref{eq:para:massive} becomes
\begin{align} \label{eq:kAB34}
k_{AB}^{34}(a,b,y,t_{kl})&= \frac{16\beta_t^2y^2}{\big[(1+y)^2-(1-y)^2 \beta_t^2\big]^2} \;
k_{AB}^{12}(a,b,y,t_{kl}) \,. 
\end{align}
The kernel can thus be related to that of the massless dipole up to a factor that encodes the dependence on the top-quark kinematics, and which is the same factor we encountered at NLO in \eqref{eq:me:nlo:massive}.

The same is true for the mixed massless-massive dipoles, for which we obtain e.g.~in the parameterisation \eqref{eq:para:mixed} for the (13) dipole,
\begin{align} \label{eq:kAB13}
k_{AB}^{13}(a,b,y,t_{kl})&=  \frac{(1-\beta_t \cos \vartheta)^4}{\big[(1-\beta_t\cos\vartheta)^2+y\,(1-\beta_t^2)\big]^2} \;
k_{AB}^{12}(a,b,y,t_{kl}) \,. 
\end{align}
whereas the kernel $k_{CD}^{13}$ follows by inverting $y\to 1/y$. The $y$-dependence is, in fact, entirely captured by the prefactor in \eqref{eq:kAB13}, since the expression for $k_{AB}^{12}$ in \eqref{eq:kAB12} turns out to be independent of $y$. We can therefore draw the same conclusions about the collinear divergences as at NLO here: the (13) dipole develops a divergence for $y\to 0$ in regions $A, B$, but not in regions $C,D$ where this limit reflects collinearity to the massive quark. 

The simple relation between the massless and massive kernels does not carry over to the gluonic channel, for which the matrix element receives non-trivial mass corrections~\cite{Czakon:2011ve}. Notice also that the expressions in \eqref{eq:kAB34} and \eqref{eq:kAB13} simplify to the massless result in \eqref{eq:kAB12} in the ultra-relativistic limit $\beta_t\to1$, which is of course also true for the more complicated gluon channel. 

For the 0-jettiness calculation in section \ref{sec:results}, one in addition needs the explicit form of the measurement function \eqref{eq:2emMeas} in the chosen parameterisations. Specifically for \mbox{region $A$}, they read
\begin{align} \label{eq:0-jettiness:M2}
& F_A^{(12)}(a,b,y,t_{kl},t_{l},t'_{5}) = 
\min\Big( \frac{b}{a + b}, \frac{a b}{(1 + a b) y} \Big)
+ \min\Big( \frac{a}{a + b}, \frac{1}{(1 + a b) y} \Big)
\,,\nonumber \\
& F_A^{(13)}(a,b,y,t_{kl},t_{l},t'_{5})  \nonumber \\
& \quad =\min\Big(\frac{b}{a+ b},\frac{ab}{(1+ab)y}+ \frac{b}{a+ b}\,\frac{\beta_t^2 \sin^2\vartheta}{(1-\beta_t\cos\vartheta)^2}-\frac{\rho\, b\, (1-2 t_k)}{\sqrt{y}}\,\frac{2\beta_t\sin\vartheta}{(1-\beta_t\cos\vartheta)}
\Big) \nonumber \\
& \quad+\min\Big(\frac{a}{a+ b},\frac{1}{(1+ab)y}+\frac{a}{a+ b}
\,\frac{\beta_t^2 \sin^2\vartheta}{(1-\beta_t\cos\vartheta)^2}
-\frac{\rho\, (1-2 t_l)}{\sqrt{y}}\,\frac{2\beta_t\sin\vartheta}{(1-\beta_t\cos\vartheta)}\,\Big) \,,\nonumber \\
& F_A^{(34)}(a,b,y,t_{kl},t_{l},t'_{5})  \nonumber \\
&\quad=\min\Big(\frac{a b}{(1+a b)y} \, \sin^2\frac{\vartheta}{2}+\frac{b}{a+b}\, \cos^2\frac{\vartheta}{2}+ \frac{\rho\, b \,(1-2t_k)}{\sqrt{y}}\, \sin\vartheta, \nonumber \\
&\quad \hspace{15mm} \frac{a b}{(1+a b)y} \, \cos^2\frac{\vartheta}{2}+\frac{b}{a+b}\, \sin^2\frac{\vartheta}{2}- \frac{\rho\, b \,(1-2t_k)}{\sqrt{y}}\, \sin\vartheta \Big)
\nonumber \\
&\quad+\min\Big(\frac{1}{(1+a b)y} \, \sin^2\frac{\vartheta}{2}+\frac{a}{a+b}\, \cos^2\frac{\vartheta}{2}+ \frac{\rho\, (1-2t_l)}{\sqrt{y}}\, \sin\vartheta, \nonumber \\
& \quad\hspace{15mm} \frac{1}{(1+a b)y} \, \cos^2\frac{\vartheta}{2}+\frac{a}{a+b}\, \sin^2\frac{\vartheta}{2}- \frac{\rho\, (1-2t_l)}{\sqrt{y}}\, \sin\vartheta \Big)\,,
\end{align}
where we introduced $\rho=\sqrt{ a/(a+b)/(1+a b)}$, and $t_{k}= t_l + t_{kl} - 2 t_{l} t_{kl} + 2\sqrt{t_{l}\bar t_{l}t_{kl}\bar t_{kl}}\,(1-t_5')$. The expressions in the other regions then follow by inverting either $y\to 1/y$ or $b\to 1/b$ (or both) as described in~\cite{Bell:2018oqa}.

Before we turn to the renormalisation of the bare soft function, let us briefly comment on our numerical implementation of the formalism. To validate our results, we implemented the above master formula in two independent codes: a {\tt SoftSERVE} extension that uses the preferred parameterisations in the frames described in section~\ref{sec:frames}, and a {\tt pySecDec}~\cite{Borowka:2017idc} implementation that employs parameterisations in the partonic centre-of-mass frame throughout. While our final numbers were produced using {\tt SoftSERVE}, it is interesting to compare its computational performance with the public version for massless soft functions. On the one hand, it is clear that the integration kernels are significantly more involved for massive soft functions, but on the other hand the matrix elements are less divergent in this case, which reduces the complexity of the integrands in the $\epsilon$-expansion. In practice, we find that the latter reduction in complexity outpaces the more complicated form of the matrix elements.  As a result, the  massive-massive dipole calculation is significantly faster than the mixed massless-massive dipole, which in turn is only marginally slower than the purely massless version of \texttt{SoftSERVE}.  \footnote{As a ballpark estimate for this comparison: to achieve percent-level accuracy on a standard laptop, the mixed dipoles with generic kinematics run for slightly less than three minutes, while the massive dipole for the same kinematics takes 45 seconds, and the massless calculation takes a bit more than two minutes.}

\section{Renormalisation}
\label{sec:ren}

In this section we discuss the renormalisation of the bare soft function in the combined SCET+HQET framework. In what follows, for simplicity, we display the $\tau$ and $\mu$ dependence, where present, only for all-order quantities; for the perturbative coefficients it is left implicit, since it can be inferred either from the defining equations or from their explicit expressions. 

We first introduce the renormalised soft function $\mathbf{S}_R(\tau,\mu)$ in terms of the bare soft function $\mathbf{S}_0(\tau)$ as
\begin{align}\label{eq:ren}
    \mathbf{S}_R(\tau,\mu) = \mathbf{Z}^\dagger(\tau,\mu) \, \mathbf{S}_0(\tau)\,  \mathbf{Z}(\tau,\mu) = \Big(e^{\ln\mathbf{Z}(\tau,\mu)}\Big)^\dagger\, e^{\ln\mathbf{S}_0(\tau)} \,e^{\ln \mathbf{Z}(\tau,\mu)} \,, 
\end{align}
where in the last equality we have re-expressed the renormalisation operator $\mathbf{Z}(\tau,\mu)$ and the bare soft function in terms of their logarithms. Starting from this expression, applying the Baker-Campbell-Hausdorff formula
\begin{align}
    e^{\mathbf{x}} e^{\mathbf{y}} e^{\mathbf{z}}\! = \! e^{(\mathbf{x+y+z})+\frac{1}{2}\big(\mathbf{
    [y,z]+[x,y]+[x,z]
    }\big)+\ldots}\; , 
\end{align}
and expanding in powers of the renormalised strong coupling $\alpha_s$,
\begin{align}
\mathbf{Z}(\tau,\mu) &= \mathbf{1} + \bigg(\frac{\alpha_s}{4 \pi}\bigg) \, \mathbf{Z}^{(1)}+\bigg(\frac{\alpha_s}{4 \pi}\bigg)^2 \,\mathbf{Z}^{(2)} + \mathcal{O}(\alpha_s^3)\, , \nonumber \\
\mathbf{S}_0 (\tau) &= \mathbf{1} + \bigg(\frac{\alpha_s}{4 \pi}\bigg) \, \mathbf{S}^{(1)}_0+\bigg(\frac{\alpha_s}{4 \pi}\bigg)^2 \,\mathbf{S}^{(2)}_0 + \mathcal{O}(\alpha_s^3)\, , 
\end{align}
one obtains the following expression for the logarithm of the renormalised soft function up to two-loop order
\begin{align}\label{eq:softren}
    \ln \mathbf{S}_R(\tau,\mu) = & \bigg( \frac{\alpha_s}{4 \pi}\bigg) \bigg(\mathbf{S}_0^{(1)} + \mathbf{Z}^{(1)} + \mathbf{Z}^{\dagger \, (1)}\bigg)+ \bigg(\frac{\alpha_s}{4 \pi}\bigg)^2 \bigg[ \bigl(\ln \mathbf{S}_0\bigr)^{(2)} + \bigl(\ln\mathbf{Z}\bigr)^{(2)} + \bigl(\ln \mathbf{Z}^{\dagger}\bigr)^{(2)} \, \nonumber \\
    & +\frac{1}{2}\bigg(\big[\mathbf{S}^{(1)}_0, \big(\mathbf{Z}^{(1)} - \mathbf{Z}^{\dagger\, (1)}\big)\big] +\big[\mathbf{Z}^{\dagger\, (1)}, \mathbf{Z}^{(1)}\big]\bigg)\bigg] + \mathcal{O}(\alpha_s^3)\, ,
\end{align}
where
\begin{align}
\mathbf{S}_0^{(1)} &= (\mu^2 \bar{\tau}^2)^{\epsilon}\, \mathbf{S}^{(1)}(\epsilon)\, ,\nonumber \\
\mathbf{S}_0^{(2)} &= (\mu^2 \bar{\tau}^2)^{2 \epsilon}\, \mathbf{S}^{(2)}(\epsilon) - \frac{\beta_0}{\epsilon} (\mu^2 \bar{\tau}^2)^{\epsilon}\, \mathbf{S}^{(1)}(\epsilon) \, 
\end{align}
are the soft-function coefficients after $\alpha_s$ renormalisation, and we defined the second-order perturbative coefficient of the logarithm of a colour operator $\mathbf{A}\in \{\mathbf{S}_0,\mathbf{Z}\}$ as
\begin{align}
    (\ln \mathbf{A})^{(2)} \equiv \mathbf{A}^{(2)} - \frac{\big(\mathbf{A}^{(1)}\big)^2}{2}\, . 
\end{align}
The renormalised soft function can likewise be expanded in powers of $\alpha_s$,
\begin{align}
\mathbf{S}_R (\tau,\mu) &= \mathbf{1} 
+ \bigg(\frac{\alpha_s}{4 \pi}\bigg) \, \mathbf{S}^{(1)}_R 
+ \bigg(\frac{\alpha_s}{4 \pi}\bigg)^2 \,\mathbf{S}^{(2)}_R 
+ \mathcal{O}(\alpha_s^3) \, , 
\end{align}
so that the last line of \eqref{eq:softren} can be re-expressed in terms of the 
one-loop renormalised soft function $\mathbf{S}^{(1)}_R=\mathbf{S}^{(1)}_0 + \mathbf{Z}^{(1)}+\mathbf{Z}^{\dagger\, (1)}$
as
\begin{align}\label{eq:nonabelian}
\frac{1}{2}\bigg(\big[\mathbf{S}^{(1)}_0, \big(\mathbf{Z}^{(1)} - \mathbf{Z}^{\dagger\, (1)}\big)\big] +\big[\mathbf{Z}^{\dagger\, (1)}, \mathbf{Z}^{(1)}\big]\bigg) \!=\! \frac{1}{2}\bigg(\big[\mathbf{S}^{(1)}_R,\big(\mathbf{Z}^{(1)} - \mathbf{Z}^{\dagger\,  (1)}\big)\big]+\big[\mathbf{Z}^{(1)},\mathbf{Z}^{\dagger\, (1)}\big]\bigg)\, .  
\end{align}
As far as the logarithm of the soft function is concerned, it is sufficient to focus on the renormalisation of the dipole and tripole correlation terms. The dipole contributions are fully renormalised by the dipole terms of the renormalisation matrices contained in $(\ln \mathbf{Z})^{(2)}$ and $(\ln \mathbf{Z}^{\dagger})^{(2)}$. The tripole terms, on the other hand, are renormalised by the explicit tripole colour correlators in $(\ln \mathbf{Z})^{(2)}$ and $(\ln \mathbf{Z}^{\dagger})^{(2)}$, together with the commutators appearing in the last line of \eqref{eq:softren}. In what follows, we discuss the renormalisation procedure for generic SCET$_\mathrm{I}$ observables (with parameter $n\neq 0$) in detail.

\subsection{Renormalised soft function}

We begin with the renormalisation-group equation (RGE) satisfied by the renormalised soft function, 
\begin{align}\label{eq:rge}
    \frac{\mathrm{d}}{\mathrm{d} \ln \mu} \mathbf{S}_R(\tau,\mu) = \mathbf{\Gamma}^{\dagger}_s(\tau,\mu) \, \mathbf{S}_R(\tau,\mu) + \mathbf{S}_R(\tau,\mu)\, \mathbf{\Gamma}_s(\tau,\mu) \, , 
\end{align}
which is governed by a soft anomalous dimension $\mathbf{\Gamma}_s(\tau,\mu)$ that is related to the renormalisation operator $\mathbf{Z}(\tau,\mu)$  via
\begin{align}
    \mathbf{\Gamma}_s(\tau,\mu) =  \mathbf{Z}^{-1}(\tau,\mu) \, \frac{\mathrm{d}}{\mathrm{d} \ln \mu} \mathbf{Z}(\tau,\mu)\,.  
\end{align}
The soft anomalous dimension can be explicitly calculated, or extracted by means of consistency relations, from the two-loop hard anomalous dimension computed in \cite{Becher:2009kw,Ferroglia:2009ii} and the appropriate (observable-dependent) beam and jet anomalous dimensions~\cite{Bell:2024lwy}.
Exploiting colour conservation to rewrite the monopole terms in the hard and collinear anomalous dimensions as dipole contributions, we obtain the following expression for the soft anomalous dimension relevant for processes with two massive quarks and an arbitrary number of massless partons
\begin{align}\label{eq:softandim}
\mathbf{\Gamma}_s(\tau,\mu) = & \sum_{(i,j)} \frac{\mathbf{T}_i \cdot \mathbf{T}_j}{2} \bigg[- \frac{\Gamma_{\rm cusp}(\alpha_s)}{n} \, L + \Gamma_{\rm cusp}(\alpha_s) \, \ln \Bigl(- \sigma_{ij}\frac{n_i\cdot n_j}{2}-i 0^+\Bigr) + \frac{\gamma^{S,\,ij}(\alpha_s)}{n}\bigg] \,\nonumber \\
& + \sum_{i,J} \mathbf{T}_i \cdot \mathbf{T}_J \bigg[-\frac{\Gamma_{\rm cusp}(\alpha_s)}{2 n} L + \, \Gamma_{\rm cusp}(\alpha_s)\, \ln \Bigl(- \sigma_{Ji}\, v_J\cdot n_i-i 0^+\Bigr) + \frac{\gamma^{S,\, iJ}(\alpha_s)}{n} \bigg] \, \nonumber\\
& + \sum_{(I,J)} \frac{\mathbf{T}_I \cdot \mathbf{T}_J}{2} \bigg[\Gamma_{\rm cusp}(\beta_{IJ},\alpha_s) +  \gamma^{S,\, I J}(\alpha_s) \bigg] \, \nonumber \\
& - \sum_{(I,J)} \sum_k \,i f^{abc}\, \mathbf{T}^a_I \mathbf{T}^b_J \mathbf{T}^c_k \,\, f_2\bigg(\beta_{IJ}, \ln \frac{-\sigma_{Jk} \, v_J\cdot n_k}{-\sigma_{Ik} \, v_I\cdot n_k}\bigg) + \mathcal{O}(\alpha_s^3)\, , 
\end{align}
where $L=-\ln(\mu^2 \bar{\tau}^2)$, and $\sigma_{\alpha\beta}$ is a sign factor with $\sigma_{\alpha\beta}=+1$ when both external partons are either incoming or outgoing, and $\sigma_{\alpha \beta}=-1$ otherwise. The notation $(\cdot,\cdot)$ in the sums  denotes unordered pairs of distinct partons. We furthermore  adopted the notation introduced in~\cite{Bell:2018vaa,Bell:2018oqa,Bell:2023yso} for the massless non-cusp anomalous dimensions that factors out the observable-specific parameter $n$. Notice also that we distinguish between the light-like and angle-dependent cusp anomalous dimensions only via their arguments in our notation. Finally, the function $f_2$ that arises in the tripole contribution can be written in the form
\begin{align} \label{eq:f2}
f_2\!\left(\beta_{IJ},\, \ln \frac{-\sigma_{Jk}\, v_J \cdot n_k}{-\sigma_{Ik}\, v_I \cdot n_k}\right) = -\frac{\alpha_s}{4\pi}\, g(\beta_{IJ})\, \Gamma_{\mathrm{cusp}}(\alpha_s)\, \ln \frac{-\sigma_{Jk}\, v_J \cdot n_k}{-\sigma_{Ik}\, v_I \cdot n_k} \,,  
\end{align}
where the function $g(\beta_{IJ})$ was first introduced and computed in~\cite{Ferroglia:2009ii}, and its explicit expression can be found in~\eqref{eq:gbeta}. We furthermore adopt the conventions of that paper, defining $\cosh(\beta_{IJ}) = - \sigma_{IJ} \,v_I \cdot v_J - i 0^+$, with the cusp angle $\beta_{IJ}$ generating a positive imaginary part for two outgoing massive partons.

Expanding the anomalous dimensions in units of $\alpha_s/(4 \pi)$, writing e.g.~$\Gamma_{\rm cusp} (\alpha_s)= \sum_{n=0}^{\infty} \Gamma_n \,( \frac{\alpha_s}{4 \pi})^{n+1}$, we solve the RGE~\eqref{eq:rge} with this soft anomalous dimension to two-loop order for $t\bar{t}$ production in hadronic collisions.~\footnote{Contributions involving both a top quark and a light parton in the final state are therefore not considered in the subsequent formulas.} After removing the NLO-squared contributions arising from exponentiation, we obtain the following expression for the logarithm of the renormalised soft function,
\begin{align}\label{eq:softrge}
   \ln \mathbf{S}_{R}(\tau,\mu) =&   \sum_{i<j} \mathbf{T}_i\cdot \mathbf{T}_j\, \bigg\{ \bigg(\frac{\alpha_s}{4 \pi}\bigg) \,\bigg[\frac{\Gamma_0}{2 n} \, L^2 - \Gamma_0 \, \ln \Big(\frac{n_i\cdot n_j}{2}\Big) L -\frac{\gamma^{S,\, ij}_0}{n} L + c^{(1)}_{ij} \bigg]  \, \nonumber \\
   & \quad \quad \quad  + \bigg(\frac{\alpha_s}{4 \pi}\bigg)^2 \bigg[  -\frac{\beta_0\Gamma_0}{6 n} \, L^3  +\beta_0\Gamma_0 \, \ln \Big(\frac{n_i\cdot n_j}{2}\Big) \frac{L^2}{2} +\frac{\beta_0\gamma^{S,\, ij}_0}{n} \frac{L^2}{2} \nonumber \\
    &\qquad \qquad + \frac{\Gamma_1}{2 n} \, L^2 - \Gamma_1 \, \ln \Big(\frac{n_i\cdot n_j}{2}\Big) L - \bigg(\frac{\gamma^{S,\,ij}_1}{n}+\beta_0 \, c^{(1)}_{ij} \bigg) L + c^{(2)}_{ij} \bigg] \bigg\}\, \nonumber \\
    & + \sum_{i,J} \mathbf{T}_i\cdot \mathbf{T}_J \, \bigg\{\bigg(\frac{\alpha_s}{4 \pi}\bigg) \,\bigg[ \frac{\Gamma_0}{4 n} L^2 - \, \Gamma_0\, \ln \big(v_{J}\cdot n_i\big) \, L - \frac{\gamma^{S,\, iJ}_0}{n} \, L + c^{(1)}_{iJ}  \bigg]  \nonumber \\
    & \qquad \quad + \bigg(\frac{\alpha_s}{4 \pi}\bigg)^2 \bigg[  -\frac{\beta_0\Gamma_0}{12 n} L^3 +\beta_0 \Gamma_0\, \ln \big(v_{J} \cdot n_i\big) \, \frac{L^2}{2} + \frac{\beta_0\gamma^{S,\, iJ}_0}{n} \, \frac{L^2}{2} \nonumber \\
    & \qquad \qquad +\frac{\Gamma_1}{4 n} L^2 -  \Gamma_1\, \ln \big(v_{J}\cdot n_i\big) \, L - \bigg(\frac{\gamma^{S,\, iJ}_1}{n} + \beta_0 \, c^{(1)}_{iJ} \bigg)\, L +  c^{(2)}_{iJ} \bigg]    \bigg\}\, \nonumber\\
    & +  \sum_{I<J} \mathbf{T}_I\cdot \mathbf{T}_J\,\bigg\{ \bigg(\frac{\alpha_s}{4 \pi}\bigg) \bigg[ -  \big(\mathrm{Re}[\Gamma_0(\beta_{IJ})]\,  +  \gamma^{S,\, I J}_0 \big)\, L + c^{(1)}_{IJ} \bigg]  \nonumber \\
    & \qquad \quad + \bigg(\frac{\alpha_s}{4 \pi}\bigg)^2 \bigg[  \beta_0 \big(\mathrm{Re}[\Gamma_0(\beta_{IJ})]\,  +  \gamma^{S,\, I J}_0 \big)\, \frac{L^2}{2}  \nonumber \\
    & \qquad \qquad -\big(\mathrm{Re}[\Gamma_1(\beta_{IJ})] +  \gamma^{S,\, I J}_1 +\beta_0 \, c^{(1)}_{IJ}\big) \, L + c^{(2)}_{IJ} \bigg] \bigg\}\, \nonumber \\
    & + \bigg(\frac{\alpha_s}{4 \pi}\bigg)^2 \Bigg\{
    \,-\sum_{k\neq l, J} \, f^{abc}  \, \mathbf{T}^a_k \mathbf{T}^b_l \mathbf{T}^c_J\, \bigg[ \frac{\pi}{2}  \,\Gamma_0 \,
    \nonumber \\
    &\qquad \qquad \times \bigg( -\Gamma_0\, \ln \big(v_{J} \cdot n_l \big) \, \frac{L^2}{2} - \frac{\gamma^{S,\, lJ}_0}{2 n} \,L^2  + c^{(1)}_{lJ}  L \bigg) + c^{(2)}_{klJ}\bigg]\, \nonumber \\
    & +    \,\sum_{K\neq I, j} \, f^{abc}  \, \mathbf{T}^a_K \mathbf{T}^b_I \mathbf{T}^c_j\, \bigg[  \frac{1}{2}  \, \mathrm{Im}[\Gamma_0(\beta_{KI})] \, 
    \nonumber \\
    &\qquad \qquad \times \bigg( -\Gamma_0\, \ln \big(v_{I} \cdot n_j \big)\, \frac{L^2}{2} - \frac{\gamma^{S,\, jI}_0}{2 n} \, L^2  + c^{(1)}_{jI} L \bigg) \,\nonumber \\
    & \qquad \qquad \quad + \, \mathrm{Im}[g(\beta_{KI})]\, 
    \Gamma_0\, \ln \bigg(\frac{ v_I\cdot n_j}{v_K\cdot n_j}\bigg) \, L+ c^{(2)}_{KIj}\bigg]\, \Bigg\}+ \mathcal{O}(\alpha_s^3)\, .
\end{align}
In the tripole contributions, one may have expected cubic $L^3$-contributions after integration, which cancel however in the sums because their leg-independent coefficients multiply the antisymmetric colour structure $f^{abc}$. It is likewise possible to show that, in our case, the following tripole combinations that appear in \eqref{eq:softrge} and include the non-cusp soft anomalous dimensions
\begin{align} \label{eq:cancelnoncusp1}
& \sum_{k\neq l, J} \, f^{abc} \, \mathbf{T}^a_k \mathbf{T}^b_l \mathbf{T}^c_J\,  \bigg[  \frac{\pi}{2}  \,\Gamma_0 \,
\frac{\gamma^{S,\, lJ}_0}{2 n} L^2 \bigg]  \nonumber \\
& \hspace{10mm}=f^{abc} \, \mathbf{T}^a_1 \mathbf{T}^b_2 \mathbf{T}^c_3 \, \frac{\pi\Gamma_0}{4 n}   \,  \big(  \gamma^{S,2 3}_0 - \gamma^{S,1 3}_0 - \gamma^{S,2 4}_0 +\gamma^{S,1 4}_0  \big) \,L^2= 0\, , \nonumber \\
& \sum_{K\neq I, j} \, f^{abc}  \, \mathbf{T}^a_K \mathbf{T}^b_I \mathbf{T}^c_j\,  \bigg[-\frac{1}{2}  \, \mathrm{Im}[\Gamma_0(\beta_{KI})] \,
\frac{\gamma^{S,\, jI}_0}{2 n} L^2\bigg]  \nonumber \\
& \hspace{10mm} = - f^{abc} \, \mathbf{T}^a_1 \mathbf{T}^b_2 \mathbf{T}^c_3 \,\frac{1}{4 n}  \, \mathrm{Im}[\Gamma_0(\beta_{34})]  \, \big( \gamma^{S,2 3}_0 - \gamma^{S,1 3}_0 - \gamma^{S,2 4}_0 +\gamma^{S,1 4}_0 \big) \, L^2 = 0\, , 
\end{align}
must cancel for any observable involving two massive particles that obeys consistency relations among anomalous dimensions.

The derivation of the solution~\eqref{eq:softrge} requires the evaluation of various colour commutators, like e.g.
\begin{align}\label{eq:comm1}
  \bigg[\sum_{i<j} \mathbf{T}_i \cdot \mathbf{T}_j \, f_{ij},\sum_{k,I} \mathbf{T}_k \cdot \mathbf{T}_I\, g_{kI}\bigg] &= \sum_{i \neq j,I} \big[\mathbf{T}_i \cdot \mathbf{T}_j,\mathbf{T}_j \cdot \mathbf{T}_I\big] \, f_{ij}\,g_{jI} \nonumber\\
& = \sum_{i \neq j,I} (-i) f^{abc} \, \mathbf{T}^a_i \mathbf{T}^b_j \mathbf{T}^c_I \, f_{ij}\,g_{jI} \, , \nonumber\\
 \bigg[\sum_{I<J} \mathbf{T}_I \cdot \mathbf{T}_J \, f_{IJ},\sum_{i,K} \mathbf{T}_i \cdot \mathbf{T}_K \, g_{iK}\bigg]  &= \sum_{I \neq J,i} \big[\mathbf{T}_I \cdot \mathbf{T}_J,\mathbf{T}_J \cdot \mathbf{T}_i\big] \, \, f_{IJ}\, g_{Ji} \nonumber\\
 &= \sum_{I \neq J,i} (-i) f^{abc} \, \mathbf{T}^a_I \mathbf{T}^b_J \mathbf{T}^c_i \, f_{IJ}\, g_{Ji} \, ,
\end{align}
where $f_{\alpha\beta}$ and $g_{\alpha\beta}$ denote generic (symmetric) dipole coefficients. In principle, for the general case with an arbitrary number of massless and massive external legs, there would also be non-vanishing contributions from the commutator $\big[\sum_{i,I} \mathbf{T}_i \cdot \mathbf{T}_I \, f_{iI} ,\sum_{j,J} \mathbf{T}_j \cdot \mathbf{T}_J \, g_{jJ}\big]$. However, for $t \bar{t}$ production considered here, this term vanishes because the logarithm $\ln (- \sigma_{Ji}\, v_J\cdot n_i-i 0^+)$ appearing in the anomalous dimension does not develop an imaginary part that could compensate for the imaginary parts generated by the commutator of the colour generators. This follows from the fact that the $t\bar{t}$ process features massless initial-state particles and massive final-state particles, so that $(\sigma_{Ji}\, v_J\cdot n_i) < 0$ and $\ln (- \sigma_{Ji}\, v_J\cdot n_i)$ is purely real. For the general case, one would also have to consider the commutators $\big[\sum_{i<j} \mathbf{T}_i \cdot \mathbf{T}_j \, f_{ij}, \sum_{l<k} \mathbf{T}_l \cdot \mathbf{T}_k \, g_{lk}\big]$ and $\big[\sum_{I<J} \mathbf{T}_I \cdot \mathbf{T}_J \, f_{IJ},\sum_{L<K} \mathbf{T}_L \cdot \mathbf{T}_K \, g_{LK} \big]$, which again vanish in the considered case, since there are at most two massless and two massive external particles on which the colour generators can act.

Notice also that in \eqref{eq:softrge} we deviate from the notation established in~\cite{Bell:2023yso} in several respects. First of all, the non-logarithmic coefficients $c^{(1,2)}_{\alpha\beta}$ in the dipole contributions are defined here by factoring out logarithms in the factorisation scale, whereas in~\cite{Bell:2023yso} we included kinematic information in the logarithms. In addition, we now work with ordered tuples in the purely massless and massive contributions, whereas we used unordered tuples in~\cite{Bell:2023yso}. Likewise, we introduced two versions of the tripole coefficients in~\cite{Bell:2023yso} -- denoted by $c^{(2)}_{ijk}$ and $\tilde{c}^{(2)}_{ijk}$ -- which again differ in whether kinematic information is absorbed into the logarithms or not. The tripole coefficients $c^{(2)}_{klJ}$ and $c^{(2)}_{KIj}$ in \eqref{eq:softrge} correspond to the tilde convention of~\cite{Bell:2023yso}, except that their normalisation differs by a factor of $(2\pi)$.

The NNLO dipole coefficients appearing in \eqref{eq:softrge} consist of two colour structures,
\begin{align} 
    c_{\alpha \beta}^{(2)} &= T_F n_f\, c_{\alpha \beta}^{(2, n_f)} + C_A\, c_{\alpha \beta}^{(2, C_A)} \, , 
\end{align}
and we present numerical results for these coefficients for 0-jettiness in section~\ref{sec:results}.  For the renormalised tripole contributions, the two sums in~\eqref{eq:softrge} can be combined and written in terms of process-dependent colour structures. For the case of four external colour-charged particles we are considering here, colour conservation implies there is only one independent tripole colour structure. Specifically, we can write the sum over the tripole coefficients as
\begin{align}\label{eq:c2tridef}
  &  -\sum_{k\neq l,J} \, f^{abc}  \, \mathbf{T}^a_k \mathbf{T}^b_l \mathbf{T}^c_J \, c^{(2)}_{kl J} + \sum_{K\neq I,j} \, f^{abc}  \, \mathbf{T}^a_K \mathbf{T}^b_I \mathbf{T}^c_j \, c^{(2)}_{KIj}
    \,\equiv\, f^{abc} \, \mathbf{T}^a_1 \mathbf{T}^b_2 \mathbf{T}^c_3 \, c^{(2)}_\textrm{tri}\, , 
\end{align}
and we present numerical results for this coefficient for 0-jettiness in section~\ref{sec:results}.

The soft function enters the factorisation and resummation formulae for 0-jettiness~\cite{Alioli:2021ggd} and for soft-gluon resummation~\cite{Ahrens:2010zv} through a colour trace with the hard function. To obtain it in this form, we evaluate the matrix elements of the colour operators appearing in \eqref{eq:softrge} and \eqref{eq:c2tridef} in the singlet--octet colour basis defined in \eqref{eq:basisvect}. We employ the normalisation conventions of~\cite{Ahrens:2010zv,Alioli:2021ggd}, in which the colour average $1/d_R^2$ over the two incoming partons is split between the hard and the soft function, each carrying a factor $1/d_R$. We therefore define
\begin{align}\label{eq:softmatrices}
    \big(\mathbf{w}_{\alpha \beta}\big)_{MN} &= \frac{1}{d_R}\langle c_M|\mathbf{T}_{\alpha}\cdot\mathbf{T}_{\beta}|c_N\rangle\, ,\nonumber \\
    \big(\mathbf{w}_{\mathrm{tri}}\big)_{MN} & = \frac{1}{d_R}\langle c_M|f^{abc}\,\mathbf{T}_1^a\mathbf{T}_2^b\mathbf{T}_3^c|c_N\rangle\, .
\end{align}
For completeness, we also define the Born soft function matrix as
\begin{align}\label{eq:softbornmatrix}
     \big(\mathbf{w}_{\mathrm{Born}}\big)_{MN} & = \frac{\langle c_M|c_N \rangle}{d_R}\, .
\end{align}
Here and above $d_R$ is the dimension of the colour representation of the incoming partons, $d_R=N$ for the $q\bar q$ and $\bar q q$ channels and $d_R=(N^2-1)$ for the $gg$ channel, with $N$ the number of colours.
The singlet--octet basis spans the space of colour-singlet states, which is two-dimensional in the former case and three-dimensional in the latter, so the soft function is a $2\times 2$ and a $3\times 3$ matrix, respectively.
In appendix~\ref{app:CB} we provide explicit results for the colour matrices defined in \eqref{eq:softmatrices} and \eqref{eq:softbornmatrix}.

\subsection{Renormalisation of the dipole contributions}

The dipole contribution to the renormalisation operators in \eqref{eq:softren} can be computed by solving the corresponding RGE
\begin{align}
    \frac{\mathrm{d}}{\mathrm{d} \ln \mu} \mathbf{Z}(\tau,\mu) = \mathbf{Z}(\tau,\mu) \, \mathbf{\Gamma}_s(\tau,\mu)   \,. 
\end{align}
The structure of the dipole part is thus entirely governed by the dipole terms of the soft anomalous dimension in \eqref{eq:softandim}. Explicitly, we find for the logarithm
\begin{align}\label{eq:lnZmatrices}
\ln \mathbf{Z}^{\mathrm{dip}}(\tau,\mu) = & -\frac{\alpha_s}{4 \pi} \bigg[\frac{\mathbf{\Gamma}^{\prime \, \mathrm{dip}}_{s,0}}{4 \epsilon^2} + \frac{{\mathbf{\Gamma}}^{\mathrm{dip}}_{s,0}}{2 \epsilon}\bigg] \, \nonumber \\
&- \bigg(\frac{\alpha_s}{4 \pi} \bigg)^2 \bigg[ -\frac{3 \beta_0 \mathbf{\Gamma}^{\prime \,\mathrm{dip}}_{s,0}}{16 \epsilon^3} + \frac{\mathbf{\Gamma}^{\prime \, \mathrm{dip}}_{s,1} - 4 \beta_0 \mathbf{\Gamma}^{\mathrm{dip}}_{s,0}}{16 \epsilon^2} + \frac{\mathbf{\Gamma}^{\mathrm{dip}}_{s,1}}{4 \epsilon}\bigg]+ \mathcal{O}(\alpha_s^3) \, , 
\end{align}
where
\begin{align}
\mathbf{\Gamma}^{\prime \, \mathrm{dip}}_s (\alpha_s)=\frac{\partial}{\partial \ln \mu } \mathbf{\Gamma}^{\mathrm{dip}}_s(\tau, \mu, \alpha_s)\, , 
\end{align}
and we have expanded the anomalous dimensions in units of $\alpha_s/(4 \pi)$ as usual. Substituting the explicit expression of the anomalous dimension \eqref{eq:softandim} into \eqref{eq:lnZmatrices}, we obtain
\begin{align}\label{eq:dipoles}
\ln \mathbf{Z}^{\mathrm{dip}}(\tau,\mu)
&=  \sum_{i < j} \mathbf{T}_i \cdot \mathbf{T}_j\, \bigg\{ -\bigg(\frac{\alpha_s}{4\pi}\bigg) \bigg[ \frac{ \Gamma_{0}}{2 n \epsilon^2} -\frac{\Gamma_{0}}{2 n \epsilon}\,L
+\frac{\Gamma_{0}}{2 \epsilon}\,\ln \Big(- \sigma_{ij}\frac{n_i\cdot n_j}{2}-i 0^+\Big)
+\frac{\gamma^{S,\,ij}_0}{2 n \epsilon}
\bigg]\nonumber \\
&   -\bigg(\frac{\alpha_s}{4\pi}\bigg)^2 \bigg[ -\frac{3 \beta_0 \Gamma_{0}}{8 n \epsilon^3} + \frac{\Gamma_{1}}{8 n \epsilon^2}+\;\frac{\beta_0\,\Gamma_{0}}{4 n \epsilon^2}\,L
-\frac{\beta_0\,\Gamma_{0}}{4 \epsilon^2}\,\ln \Big(- \sigma_{ij}\frac{n_i\cdot n_j}{2}-i 0^+\Big)\, \nonumber \\
& -\frac{\beta_0\,\gamma^{S,\,ij}_0}{4n\epsilon^2}
-\frac{\Gamma_{1}}{4n\epsilon}\,L
+\frac{\Gamma_{1}}{4\epsilon}\, \ln \Big(- \sigma_{ij}\frac{n_i\cdot n_j}{2}-i 0^+\Big)
+\frac{\gamma^{S,\,ij}_1}{4n\epsilon}
\bigg]\nonumber \bigg\} \nonumber\\
& + \sum_{i,J} \mathbf{T}_i \cdot \mathbf{T}_J \bigg\{-\bigg(\frac{\alpha_s}{4\pi}\bigg) \bigg[
\frac{\Gamma_{0}}{4 n \epsilon^2}
-\frac{\Gamma_{0}}{4 n \epsilon}\,L
+\frac{\Gamma_{0}}{2 \epsilon}\,\ln \Big(- \sigma_{Ji}\, v_J \cdot n_i - i 0^+\Big)
+\frac{\gamma^{S,\,iJ}_0}{2 n \epsilon}
\bigg] \nonumber\\
& -\bigg(\frac{\alpha_s}{4\pi}\bigg)^2\bigg[
-\frac{3 \beta_0 \Gamma_{0}}{16 n \epsilon^3}
+\frac{\Gamma_{1}}{16 n \epsilon^2}
+\frac{\beta_0 \Gamma_{0}}{8 n \epsilon^2}\,L
-\frac{\beta_0 \Gamma_{0}}{4 \epsilon^2}\, \ln \Big(- \sigma_{Ji}\, v_J \cdot n_i -i 0^+\Big) \nonumber \\
&-\frac{\beta_0 \gamma^{S,\,iJ}_0}{4 n \epsilon^2} -\frac{\Gamma_{1}}{8 n \epsilon}\,L
+\frac{\Gamma_{1}}{4 \epsilon}\, \ln \Big(- \sigma_{Ji}\, v_J \cdot n_i -i 0^+\Big)
+\frac{\gamma^{S,\,iJ}_1}{4 n \epsilon}
\bigg]\bigg\} \nonumber \\
& + \sum_{I < J} \mathbf{T}_I \cdot \mathbf{T}_J\, \bigg\{  -\bigg(\frac{\alpha_s}{4\pi}\bigg) \bigg[ \frac{\Gamma_{0}(\beta_{IJ})}{2 \epsilon} +  \frac{\gamma^{S,\, I J}_0}{2 \epsilon} \bigg]\nonumber \\
&  -\bigg(\frac{\alpha_s}{4\pi}\bigg)^2 \bigg[ - \frac{\beta_0 \Gamma_{0}(\beta_{IJ})}{4 \epsilon^2} - \frac{\beta_0  \gamma^{S,\, I J}_0}{4 \epsilon^2}  +  \frac{\Gamma_{1}(\beta_{IJ})}{4 \epsilon} +  \frac{\gamma^{S,\, I J}_1}{4 \epsilon}\bigg] \bigg\}
+ \mathcal{O}(\alpha_s^3).  
\end{align}
Note that, in order to subtract the poles of the bare soft function, one must combine the logarithm of the renormalisation operator $\mathbf{Z}(\tau,\mu)$ with its Hermitian conjugate, as prescribed by \eqref{eq:softren}. As it is clear from the structure in \eqref{eq:dipoles}, the renormalisation can be implemented for each individual dipole separately.

Turning to the specific 0-jettiness soft function that we will analyse in detail in the following section, the corresponding anomalous dimensions are all known to two-loop order, and we can thus verify if the first line of \eqref{eq:softren} removes all poles from the bare soft function for each  dipole. To this end, one can extract the non-cusp soft anomalous dimensions via consistency relations by employing the observable-independent hard anomalous dimensions collected in appendix~\ref{app:AD} along with the specific 0-jettiness non-cusp beam anomalous dimensions. The result can be expressed in terms of two independent structures,
\begin{align}
\gamma^{S,\,ij}(\alpha_s)&= 2\gamma^{S}(\alpha_s)\,,\nonumber \\
\gamma^{S,\,iJ}(\alpha_s)&= \gamma^{S}(\alpha_s) + n \overline \gamma^{S}(\alpha_s) \,, \nonumber\\
\gamma^{S,\,IJ}(\alpha_s)&= 2 \overline \gamma^{S}(\alpha_s)\, , 
\end{align}
where $n=1$ for 0-jettiness, and the respective coefficients are given by
\begin{align}
\gamma_0^S &= 0\,, \hspace{18.5mm}
\gamma_1^S =C_A \,\Big( \frac{404}{27} - \frac{11 \pi^2}{18} - 14 \zeta_3\Big) 
+ T_F n_f \,\Big(-\frac{112}{27} + \frac{2 \pi^2}{9} \Big)\,,
\nonumber\\
\overline\gamma_0^S &= -2 \,, \qquad\qquad
\overline\gamma_1^S = C_A\left(-\frac{98}{9} + \frac{2\pi^2}{3} - 4\zeta_3\right) + \frac{40}{9} \, T_F n_f\,.
\end{align}

Note that the massive contribution $\overline{\gamma}^{S}(\alpha_s)$ to the non-cusp soft anomalous dimension follows directly from the hard anomalous dimension: since massive partons do not give rise to collinear functions in this setup, no observable-dependent ingredient enters the corresponding consistency relation, and this part of the soft anomalous dimension is therefore observable independent.

\subsection{Renormalisation of the tripole contributions}

The renormalisation of the tripole contributions consists of two parts:  the explicit tripole colour correlators in $\mathbf{Z}^{(2)}$ and $\mathbf{Z}^{\dagger\,(2)}$, and the tripoles induced by the commutators in \eqref{eq:nonabelian}. We now address each of these contributions in turn.

\paragraph{Tripole correlators in $\mathbf{Z}^{(2)}$:} For processes with at least two massive external quarks, the renormalisation operator contains explicit tripole correlators starting at $\mathcal{O}(\alpha_s^2)$.  To find these, we start from the tripole contribution to the soft anomalous dimension in the last line of \eqref{eq:softandim},
\begin{align}
\mathbf{\Gamma}_s^{\mathrm{tri}}(\tau,\mu) = 
& - \sum_{(I,J)} \sum_k \,i f^{abc}\, \mathbf{T}^a_I \mathbf{T}^b_J \mathbf{T}^c_k \,\, f_2\bigg(\beta_{IJ}, \ln \frac{-\sigma_{Jk} \, v_J\cdot n_k}{-\sigma_{Ik} \, v_I\cdot n_k}\bigg) + \mathcal{O}(\alpha_s^3)\, , 
\end{align}
with the function $f_2$ given in \eqref{eq:f2}. Using colour conservation as well as the specific back-to-back kinematics of a $2\to2$ process with $p_3\cdot p_2 = p_4\cdot p_1$ and $ p_4\cdot p_2 = p_3\cdot p_1$, we find that the tripole sum for $t\bar{t}$ production can be cast into the form
\begin{align}
\mathbf{\Gamma}_s^{\mathrm{tri}}(\tau,\mu) = 
\bigg(\frac{\alpha_s}{4\pi}\bigg)^2 
i f^{abc}\, \mathbf{T}^a_1 \mathbf{T}^b_2 \mathbf{T}^c_3 \; 4\Gamma_0\, g(\beta_{34})\, \ln \bigg(\frac{ v_4 \cdot n_1}{v_3 \cdot n_1} \bigg)+ \mathcal{O}(\alpha_s^3)\, , 
\end{align}
which determines the tripole contribution to the renormalisation operator, 
\begin{align}
\mathbf{Z}^{(2),\mathrm{tri}} = 
i f^{abc}\, \mathbf{T}^a_1 \mathbf{T}^b_2 \mathbf{T}^c_3 \; \bigg(-\frac{\Gamma_0}{\epsilon} \bigg)\, g(\beta_{34})\, \ln \bigg(\frac{ v_4 \cdot n_1}{v_3 \cdot n_1} \bigg)\, . 
\end{align}
The relevant sum of the renormalisation operator and its Hermitian conjugate in the first line of \eqref{eq:softren} then projects out the imaginary part of the function $g(\beta_{34})$ in \eqref{eq:gbeta},
\begin{align} \label{eq:z2plusz2}
\mathbf{Z}^{(2),\mathrm{tri}} + \mathbf{Z}^{\dagger \, (2),\mathrm{tri}}= 
 f^{abc}\, \mathbf{T}^a_1 \mathbf{T}^b_2 \mathbf{T}^c_3 \; \bigg(\frac{2\Gamma_0}{\epsilon} \bigg)\,  \mathrm{Im}[g(\beta_{34})]\, \ln \bigg(\frac{ v_4 \cdot n_1}{v_3 \cdot n_1} \bigg)\, , 
\end{align}
with $\beta_{34} = i \pi -\ln \big((1+\beta_t)/(1-\beta_t)\big)$, which provides the required imaginary part. 

\paragraph{Tripole contributions from commutators:} The right-hand side of \eqref{eq:nonabelian} contains several commutators that also generate tripole contributions. Starting with the commutator of the renormalisation operator and its Hermitian conjugate, one can use the arguments around \eqref{eq:cancelnoncusp1} and \eqref{eq:comm1} to show that
\begin{align}
 \frac{1}{2} \big[\mathbf{Z}^{(1)},\mathbf{Z}^{\dagger\, (1)}\big] =&  \sum_{i \neq j,I}  f^{abc} \, \mathbf{T}^a_i \mathbf{T}^b_j \mathbf{T}^c_I \;  \bigg(-\frac{\pi\Gamma_{0}^2}{4 \epsilon^2 } 
\bigg) \; \ln \big( v_I \cdot n_j\big)  \,  \nonumber \\
&+ \sum_{I \neq J,i} f^{abc} \, \mathbf{T}^a_I \mathbf{T}^b_J \mathbf{T}^c_i \;  \bigg(\frac{\mathrm{Im}[\Gamma_{0}(\beta_{IJ})]\, \Gamma_{0}}{4 \epsilon^2} \bigg) \;\ln \big( v_J \cdot n_i\big)\, , 
\end{align}
for $t \bar{t}$ production. Evaluating the tripole sum explicitly using colour conservation then gives the expression
\begin{align} \label{eq:ZZd}
 \frac{1}{2} \big[\mathbf{Z}^{(1)},\mathbf{Z}^{\dagger\, (1)}\big] &=  f^{abc} \, \mathbf{T}^a_1 \mathbf{T}^b_2 \mathbf{T}^c_3 \;  \bigg(\frac{\Gamma_{0}}{2 \epsilon^2 } 
\bigg) \, \big( \mathrm{Im}[\Gamma_{0}(\beta_{34}) ]- \pi \Gamma_0 \big) \, \ln \bigg(\frac{ v_4 \cdot n_1}{v_3 \cdot n_1} \bigg)\, ,
\end{align}
in the considered back-to-back kinematics. 

In addition, the first commutator on the right-hand side of \eqref{eq:nonabelian} involves the one-loop renormalised soft function $\mathbf{S}^{(1)}_R$,  given in \eqref{eq:softrge}, which is sufficient for the purpose of removing the poles. However, the renormalisation also generates finite contributions, and to capture these, the one-loop soft function must be retained through $\mathcal{O}(\epsilon)$, since these terms multiply the $1/\epsilon$ poles of $\mathbf{Z}^{(1)}-\mathbf{Z}^{\dagger(1)}$ in \eqref{eq:nonabelian}.
Including these $\mathcal{O}(\epsilon)$ terms, we find
\begin{align}\label{eq:softeps}
    \mathbf{S}^{(1)}_R &= \sum_{i<j} \mathbf{T}_i\cdot \mathbf{T}_j\, \bigg[\frac{\Gamma_{0}}{2 n} \, L^2 -\Gamma_{0}  \, \ln \Big(\frac{n_i\cdot n_j}{2}\Big)  L - \frac{\gamma^{S,\, ij}_0}{n} L + c^{(1)}_{ij}\, \nonumber \\
    & \hspace{25mm} +\epsilon \,\bigg(-\frac{\Gamma_{0}}{6 n} \, L^3 +\frac{\Gamma_{0}}{2} \, \ln \Big(\frac{n_i\cdot n_j}{2}\Big)  L^2 +\frac{\gamma^{S,\, ij}_0}{2 n} L^2 - c^{(1)}_{ij}\, L + c^{(1,\epsilon)}_{ij} \bigg)\bigg]\, \nonumber \\
    & + \sum_{i,J} \mathbf{T}_i\cdot \mathbf{T}_J \, \bigg[ \frac{\Gamma_{0}}{4 n} L^2 - \, \Gamma_{0}\, \ln \big(v_J \cdot n_i \big)\, L - \frac{\gamma^{S,\, iJ}_0}{n} \, L + c^{(1)}_{iJ} \,\nonumber \\
    & \hspace{25mm} +\epsilon \,\bigg(-\frac{\Gamma_{0}}{12 n} L^3 + \, \frac{\Gamma_{0}}{2}\, \ln \big(v_J \cdot n_i \big)\, L^2 + \frac{\gamma^{S,\, iJ}_0}{2 n} \, L^2 - c^{(1)}_{iJ} \, L + c^{(1,\epsilon)}_{iJ} \bigg)\bigg] \, \nonumber\\
    & +  \sum_{I<J} \mathbf{T}_I\cdot \mathbf{T}_J\, \bigg[ -\big(\mathrm{Re}[\Gamma_{0}(\beta_{IJ})] \,  +  \gamma^{S,\, I J}_0\big)\, L + c^{(1)}_{IJ}\, \nonumber \\
    & \hspace{25mm} +\epsilon \,\biggl(\big(\mathrm{Re}[\Gamma_{0}(\beta_{IJ})] \,  +  \gamma^{S,\, I J}_0\big)\, \frac{L^2}{2} - c^{(1)}_{IJ} \, L + c^{(1, \epsilon)}_{IJ} \biggr)\bigg] \,, 
\end{align}
which yields the commutator
\begin{align}
& \frac{1}{2} \big[\mathbf{S}^{(1)}_{R},\big(\mathbf{Z}^{(1)} - \mathbf{Z}^{\dagger\,  (1)}\big)\big] \,  \nonumber \\
&=  \sum_{i \neq j,I}  f^{abc} \, \mathbf{T}^a_i \mathbf{T}^b_j \mathbf{T}^c_I \;  \bigg(\frac{-\pi\Gamma_{0}}{2\epsilon }  
\bigg) \,\bigg[ -\Gamma_{0}\, \ln \big(v_I \cdot n_j\big) \, L + c^{(1)}_{j I} 
\nonumber \\
& \hspace{53mm}
+\epsilon \,\bigg( \frac{\Gamma_{0}}{2}\, \ln \big(v_I \cdot n_j\big) \, L^2 - c^{(1)}_{j I} L + c^{(1,\epsilon)}_{j I}\bigg)\bigg]  \,  \nonumber \\[0.4em]
&+ \sum_{I \neq J,i} f^{abc} \, \mathbf{T}^a_I \mathbf{T}^b_J \mathbf{T}^c_i \;  \bigg(\frac{\mathrm{Im}[\Gamma_{0}(\beta_{IJ})]}{2 \epsilon} \bigg) \, 
\bigg[ -\Gamma_{0}\, \ln \big(v_J \cdot n_i\big) \, L + c^{(1)}_{i J} 
\nonumber \\
& \hspace{53mm}
+\epsilon \,\bigg( \frac{\Gamma_{0}}{2}\, \ln \big(v_J \cdot n_i\big) \, L^2 - c^{(1)}_{i J} L + c^{(1,\epsilon)}_{i J}\bigg)\bigg]  \,. 
\end{align}
Summing over the tripole colour structures then gives, using $c^{(1)}_{23}=c^{(1)}_{14}$ and $c^{(1)}_{24}=c^{(1)}_{13}$ and similar relations for the $c^{(1,\epsilon)}_{\alpha\beta}$ coefficients in the back-to-back kinematics,
\begin{align}
 &\frac{1}{2} \big[\mathbf{S}^{(1)}_{R},\big(\mathbf{Z}^{(1)} - \mathbf{Z}^{\dagger\,  (1)}\big)\big] \nonumber \\
& \quad =  f^{abc} \, \mathbf{T}^a_1 \mathbf{T}^b_2 \mathbf{T}^c_3 \;   \big( \mathrm{Im}[\Gamma_{0}(\beta_{34}) ]- \pi \Gamma_0 \big)   \bigg\{ \bigg( -\Gamma_0  \, \ln \bigg(\frac{ v_4 \cdot n_1}{v_3 \cdot n_1} \bigg)L + c^{(1)}_{14} - c^{(1)}_{13} \bigg) \frac{1}{\epsilon} \nonumber \\
& \hspace{10mm}+ \frac{\Gamma_0}{2}  \, \ln \bigg(\frac{ v_4 \cdot n_1}{v_3 \cdot n_1} \bigg)L^2 -\big( c^{(1)}_{14} - c^{(1)}_{13} \big) L +c^{(1,\epsilon)}_{14} - c^{(1,\epsilon)}_{13} \bigg\}\,. 
\label{eq:s1z1}
\end{align}


\paragraph{Total tripole contribution:} 
The complete renormalisation of the tripole contributions requires summing the expressions in \eqref{eq:z2plusz2}, \eqref{eq:ZZd} and \eqref{eq:s1z1}, in accordance with \eqref{eq:softren} and \eqref{eq:nonabelian}. As we explicitly verified for the 0-jettiness soft function, this combination precisely cancels all poles arising from the direct diagrammatic computation of the bare tripole corrections, and it yields in addition the finite contribution given in the second line of \eqref{eq:s1z1}. The logarithms in this  term then combine with similar logarithms of the bare tripoles --  generated by poles multiplying the scale-dependent prefactor $(\mu^2 \bar{\tau}^2)^{2\epsilon}$ -- and we have verified that their sum reproduces the logarithmic structure of the tripole terms in the renormalised soft function in \eqref{eq:softrge}. This provides an independent check of the solution of the RG equation.

\section{Results}
\label{sec:results}

\begin{figure}[t!]
	\centering{
		\includegraphics[height=0.22\textheight]{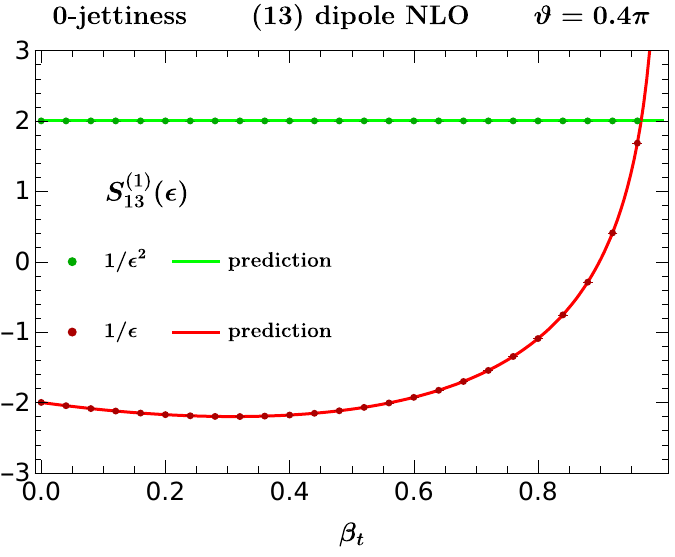}
        \hspace{10mm}
        \includegraphics[height=0.22\textheight]{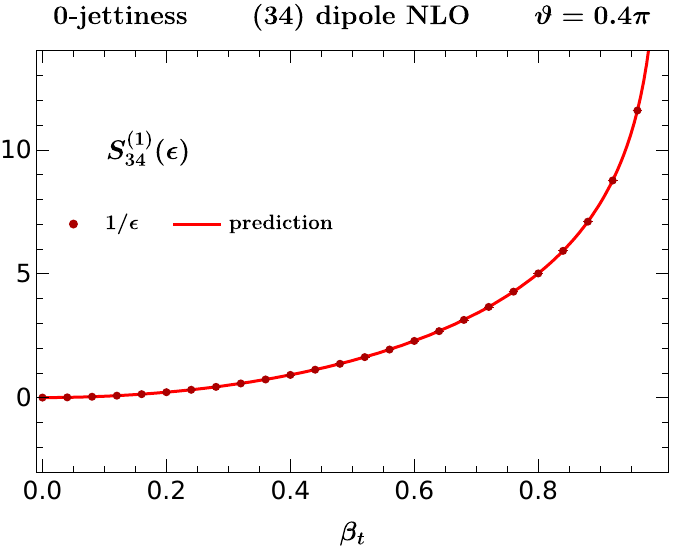}
        \\[1.0em]
        \includegraphics[height=0.22\textheight]{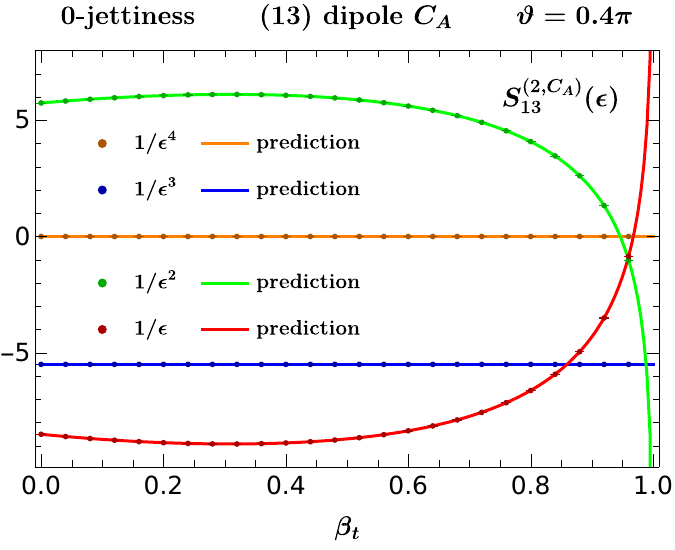}
        \hspace{10mm}
        \includegraphics[height=0.22\textheight]{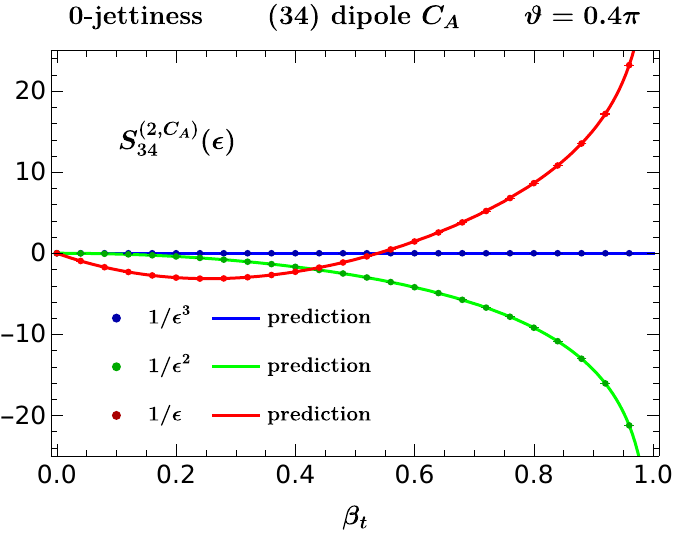}\\[1.0em]
        \includegraphics[height=0.22\textheight]{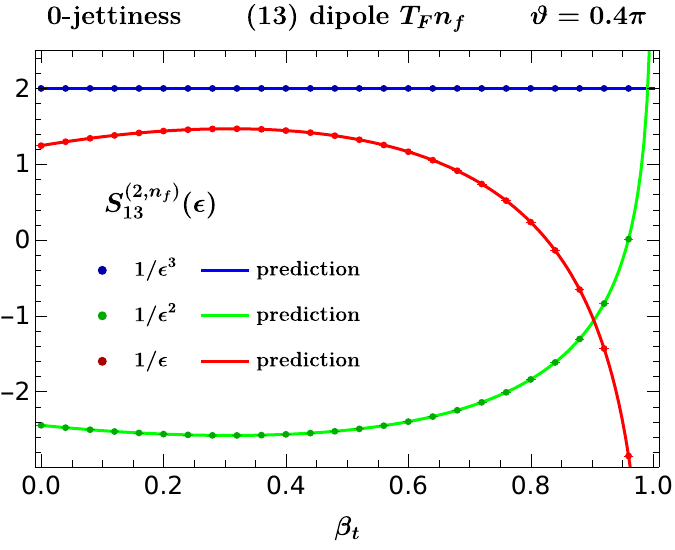}
        \hspace{10mm}
        \includegraphics[height=0.22\textheight]{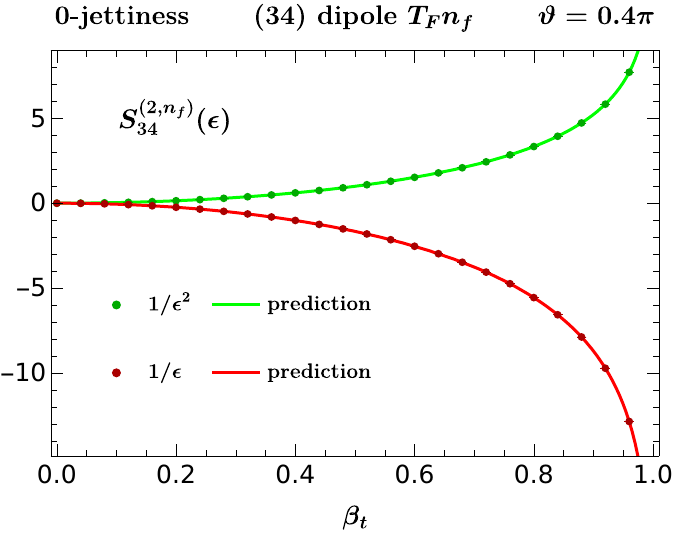}}
	\caption{Divergence structure of the dipole contributions to the bare 0-jettiness soft function. The plots show the $\beta_t$-dependence of the NLO dipoles (upper row) and the two colour structures of the NNLO dipoles (middle and lower rows) for the mixed (left column) and purely massive dipole (right column). The dots represent our numerical results and the solid lines show the RG predictions.}
	\label{fig:dipole:divergences}
\end{figure}

The formalism developed in the previous sections can be used for a wide class of SCET$_{\mathrm{I}}$ observables that comply with non-Abelian exponentiation. For a detailed discussion about the requirements on the observables, we refer to~\cite{Bell:2018oqa}. Here, we apply the framework to the 0-jettiness soft function in hadronic $t\bar{t}$ production. To this end, we implemented the master formulae from above in the {\tt SoftSERVE} distribution, setting the parameter $n=1$ and the respective measurement functions to the expressions given in \eqref{eq:0-jettiness:M1} and\eqref{eq:0-jettiness:M2}.

The soft-function coefficients depend on the top-quark velocity $\beta_t\in [0,1)$ and the scattering angle $\vartheta\in [0,\pi]$ in the partonic centre-of-mass frame. In our numerical approach, we sample this dependence in steps of $0.04$ for the velocity, and $0.04\pi$ for the scattering angle. Excluding specific points for which the result is not defined -- i.e.~the threshold limit $\beta_t=0$ for some of the tripoles -- this yields around  $600$ sampled phase-space points.

The calculation furthermore involves six dipole and one tripole colour structure. Due to the symmetries from exchanging the two massless (or massive) partons in back-to-back configuration, the mixed dipoles are, however, not independent. As we argued around \eqref{eq:NLO:symmetries}, the symmetries imply that their coefficients defined in \eqref{eq:softrge} satisfy the relations
\begin{equation}
    c_{13}^{(i)}(\beta_t, \vartheta) = c_{24}^{(i)}(\beta_t, \vartheta)= c_{14}^{(i)}(\beta_t, \pi-\vartheta)= c_{23}^{(i)}(\beta_t, \pi-\vartheta) 
\end{equation}
at arbitrary order $i$ in perturbation theory. The purely massless dipole, on the other hand, does not see the top quarks and therefore has no kinematic information. The relevant numbers can be read off from the massless 0-jettiness calculation~\cite{Kelley:2011ng,Monni:2011gb}, which in our conventions become $c_{12}^{(1)} = \pi^2$ and
\begin{align}
    c_{12}^{(2, n_f)} = -\frac{80}{81} - \frac{154 \pi^2}{27} + \frac{104 \zeta_3}{9}\,, \qquad\quad
    c_{12}^{(2, C_A)} = \frac{2140}{81} + \frac{871 \pi^2}{54} - \frac{14 \pi^4}{15} - \frac{286\zeta_3}{9} \,. 
\end{align}
In contrast, the dipoles with at least one massive leg are non-trivial, and we provide numerical grids for the two independent (13) and (34) dipoles, as well as the colour-summed tripole contribution, in the ancillary electronic files. Whereas these numbers were obtained using the procedure described in section~\ref{sec:technical} -- where we chose appropriate frames for each contribution -- we verified our results with an independent \texttt{pySecDec} implementation that uses parameterisations in the partonic centre-of-mass frame throughout. In the remainder of this section, we illustrate some features of the results based on one-dimensional projections, before we present the full two-dimensional grids.

Starting with the dipole contributions, we first compare the divergence structure of the bare soft function to the RG prediction in \eqref{eq:softren}, by scanning the top-quark velocity $\beta_t$ and keeping the scattering angle $\vartheta=0.4\pi$ fixed. For the NLO results shown in the upper panels of Fig.~\ref{fig:dipole:divergences}, our numerical results for $S^{(1)}_{\alpha\beta}(\epsilon)$ (indicated by the dots) show perfect agreement with the RG predictions (solid lines) within uncertainties that are not visible on the scale of the plots. The same is true for the NNLO results shown in the remaining panels of Fig.~\ref{fig:dipole:divergences}, where we display the coefficients of the two colour structures, $S_{\alpha\beta}^{(2,C_A)}(\epsilon)$ (middle row) and $S_{\alpha\beta}^{(2,n_f)}(\epsilon)$ (lower row), defined after coupling renormalisation as
\begin{align}
C_A\, S_{\alpha\beta}^{(2,C_A)}(\epsilon) &+ T_F n_f\, S_{\alpha\beta}^{(2,n_f)}(\epsilon) \\
&\equiv C_A\, S_{\alpha\beta}^{(2,gg)}(\epsilon)+ T_F n_f\, S_{\alpha\beta}^{(2,q\bar{q})}(\epsilon) + C_A\, S_{\alpha\beta}^{(2,Re)}(\epsilon) - \frac{\beta_0}{\epsilon}\, S_{\alpha\beta}^{(1)}(\epsilon) \, ,\nonumber
\end{align}
where the $\beta_0$-term contains a piece proportional to $C_A$ and a piece proportional to $T_F n_f$; these results are again plotted against the analytic pole predictions from the RG equations.

\begin{figure}[t!]
	\centering{
		\includegraphics[height=0.22\textheight]{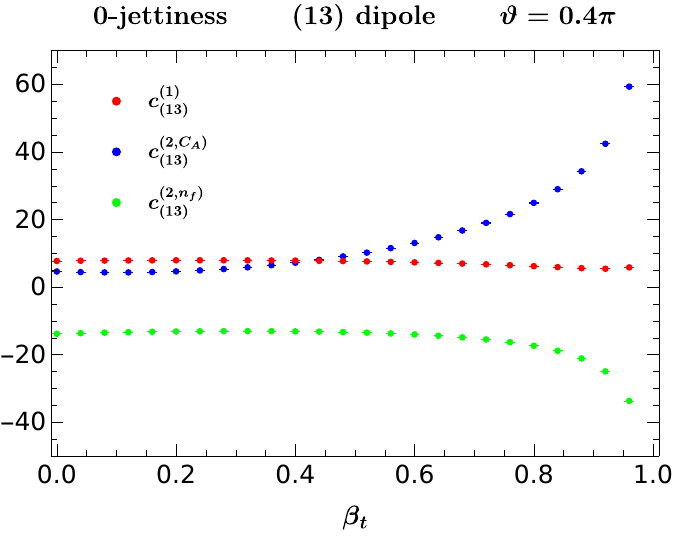}
        \hspace{10mm}
        \includegraphics[height=0.22\textheight]{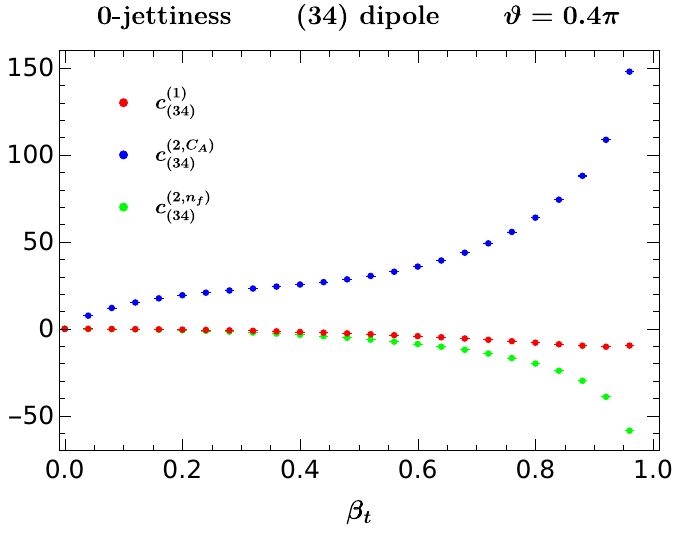}
        }
	\caption{Dipole contributions to the 0-jettiness soft function. The plots show the coefficients $c_{ij}^{(1)}$ and $c_{ij}^{(2)}=T_F n_f \,c_{ij}^{(2,n_f)}+ C_A \, c_{ij}^{(2,C_A)}$ of the renormalised soft function~\eqref{eq:softrge} for the mixed (left) and  massive dipole (right) as a function of the top-quark velocity $\beta_t$.}
	\label{fig:dipole:finite}
\end{figure}

We next turn to the finite dipole contributions $c_{\alpha\beta}^{(i)}$ in the renormalised soft function \eqref{eq:softrge}, which are shown similarly in Fig.~\ref{fig:dipole:finite} for fixed scattering angle $\vartheta=0.4\pi$. Whereas our NLO results (red dots), once combined with explicit colour matrices in a given colour basis, reproduce those of~\cite{Alioli:2021ggd}, the NNLO results (blue and green dots) are new. Once again, their numerical uncertainties are too small to be visible in the plots, but they can be reconstructed from the ancillary files.

\begin{figure}[t!]
	\centering{
	\includegraphics[height=0.17\textheight]{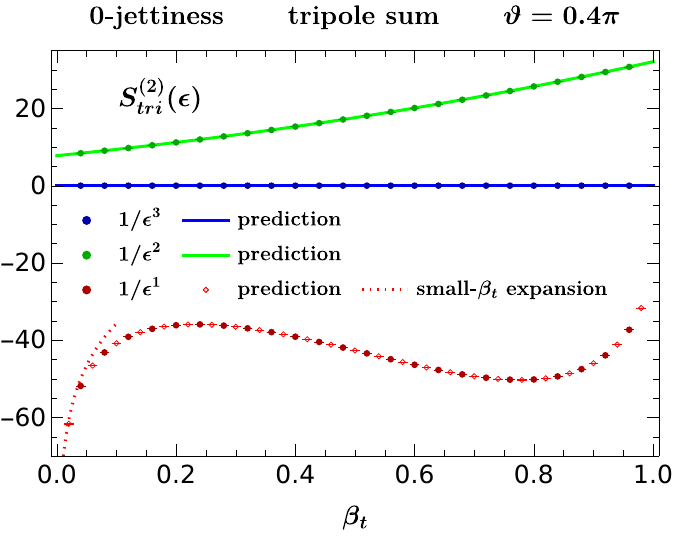}
    \includegraphics[height=0.17\textheight]{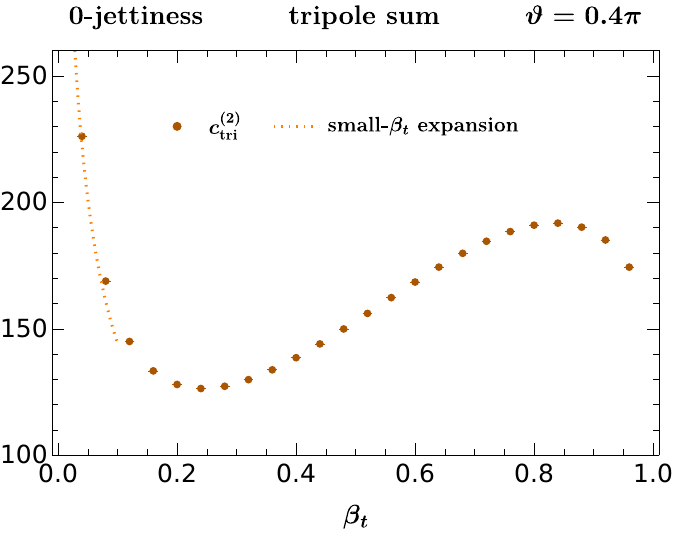}
    \includegraphics[height=0.17\textheight]{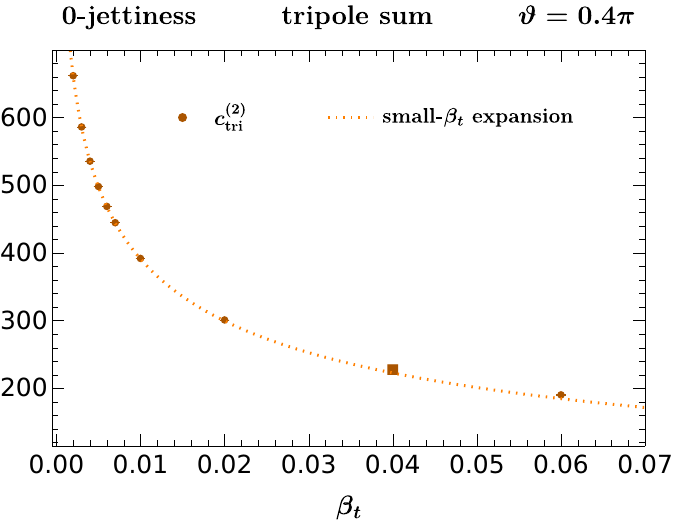}
        }
	\caption{Tripole contribution to the 0-jettiness soft function. Left: Divergences of the bare tripole sum defined in~\eqref{eq:tripolesum:bare}. Our numerical results (dots) are compared to the RG prediction (red diamonds) as well as a small-$\beta_t$ approximation for the $1/\epsilon$ coefficient (dotted line). Middle: Renormalised tripole sum in the convention of \eqref{eq:c2tridef}. The numerical results (dots) are compared against the small-$\beta_t$ expansion (dotted line). Right: The same in the small-$\beta_t$ region, where the square corresponds to the smallest value of $\beta_t$ included in the grid, i.e.~the leftmost point in the middle plot.
}
	\label{fig:tripoles}
\end{figure}

We finally consider the tripole contributions, which cannot be renormalised on the level of individual tripoles as argued before. We therefore display the poles of the bare tripole sum defined by
\begin{align} \label{eq:tripolesum:bare}
\sum_{\alpha\neq\beta\neq\gamma}f^{abc}\,\mathbf{T}^a_\alpha\mathbf{T}^b_\beta\mathbf{T}^c_\gamma \, S^{(2,\mathrm{Im})}_{\alpha\beta\gamma}(\epsilon) \equiv f^{abc}\,\mathbf{T}^a_1\mathbf{T}^b_2\mathbf{T}^c_3 \, S_{\rm tri}^{(2)}(\epsilon)
\end{align}
as a function of the top-quark velocity $\beta_t$ in the left panel of Fig.~\ref{fig:tripoles}. As some tripoles turn out to be divergent as $\beta_t\to 0$, we  carefully extracted the leading behaviour in this limit analytically. Explicitly, we find for the tripole sum
\begin{equation}
  S_\textrm{tri}^{(2)}(\epsilon)= \frac{8\pi\cos\vartheta}{\epsilon^2}+\frac{16\pi\cos\vartheta\ln \beta_t}{\epsilon}+\mathcal{O}(\epsilon^0,\beta_t)\, , 
\end{equation}
and we included the respective (divergent) $1/\epsilon$ coefficient as a dotted red line in the left panel of Fig.~\ref{fig:tripoles}. Interestingly, the Coulomb singularities associated with the loops connecting two massive legs cancel in the tripole sum, transforming the power divergence into a logarithmic one. As before, our numerical results (dots) show perfect agreement with the RG predictions (red diamonds) for all kinematic configurations.

In the middle panel of Fig.~\ref{fig:tripoles}, we present the corresponding finite term $c^{(2)}_\textrm{tri}$ of the renormalised soft function according to the conventions in \eqref{eq:c2tridef}. In the small $\beta_t$-limit, we now obtain
\begin{equation}\label{eq:tripsmallbeta}
    c_\textrm{tri}^{(2)} = 8\pi \cos\vartheta \,\big(2\ln^2 \beta_t - 4\ln^2 2+\pi^2\big) +\mathcal{O}(\beta_t),
\end{equation}
which is shown as a dotted line in the figure. Similar to the bare results, the renormalised tripole sum is thus again free of Coulomb singularities, leaving only a logarithmic divergence as $\beta_t \to 0$. The small $\beta_t$ region is also displayed in the right panel of Fig.~\ref{fig:tripoles}, for which we sampled a number of additional points (not included in the grid files) to better resolve the threshold behaviour. In fact, the red square for $\beta_t=0.04$ corresponds to the smallest value of $\beta_t$ included in the grids. As the plot suggests, the agreement between the analytic approximation and the numerical results is clearly sufficient for an accurate interpolation between the grid points and the asymptotic expansion.

\begin{figure}[t]
    \centering
    \includegraphics[height=0.23\textheight]{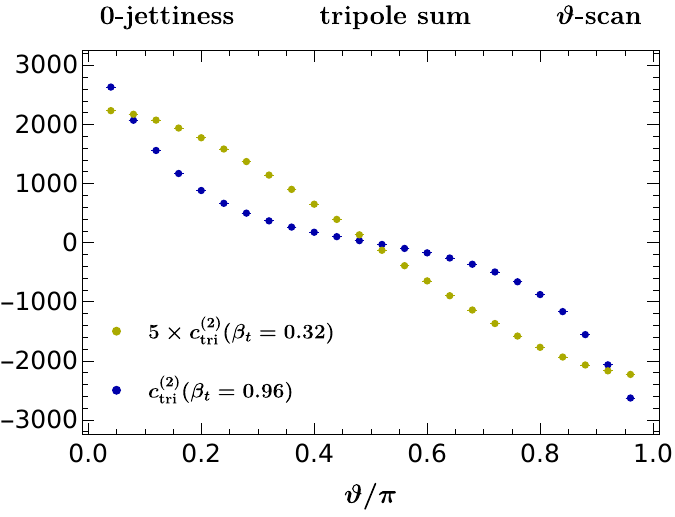}
    \hspace{10mm}
    \includegraphics[height=0.23\textheight]{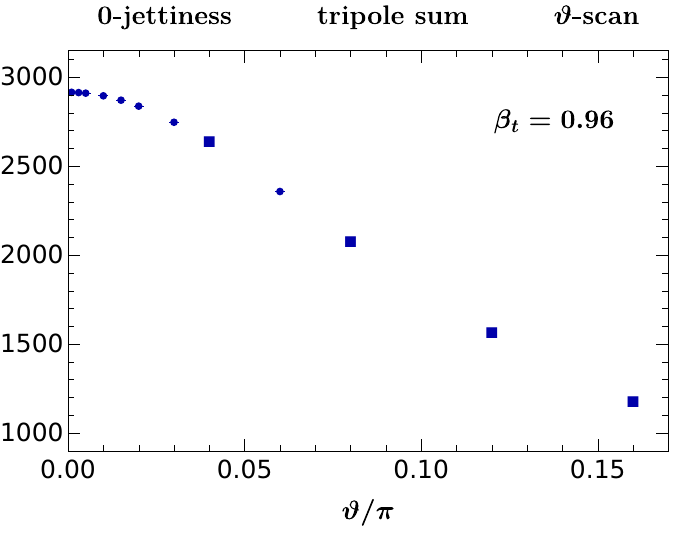}
    \caption{Left: Angular dependence of the renormalised tripole sum for two different values of the top-quark velocity $\beta_t$. Right: Zoom into the endpoint region for small \mbox{angles $\vartheta$} and $\beta_t=0.96$. The numbers included in the grids are indicated by the squares, and the additional points in the endpoint region by the circles.
    }
    \label{fig:thetascan}
\end{figure}

In Fig.~\ref{fig:thetascan} we turn the argument around and show the dependence on the scattering \mbox{angle $\vartheta$}, keeping the top-quark velocity $\beta_t$ fixed. Whereas the dipoles do not show particularly interesting features in the $\vartheta$-scans, the situation is different for the tripole contributions, which we show on the renormalised level in the left panel of Fig.~\ref{fig:thetascan}. First of all, the symmetry from exchanging two massless (or massive) legs now manifests as an anti-symmetry under $\vartheta\to\pi-\vartheta$.  In addition, we see that the tripole sum plateaus towards a finite value as $\vartheta\rightarrow 0$ for small values of $\beta_t$, whereas the tripole sum appears to be divergent in this limit for large values of $\beta_t$. In order to understand this behaviour, we calculated some additional points at very small angles (again not included in the grids), and in the right panel of Fig.~\ref{fig:thetascan} we zoom into the apparent divergence. Interestingly, this shows that the plateau is also present in these cases, and that the tripole sum alternates between concave and convex behaviour as $\vartheta$ is varied. For sufficiently small values of $\vartheta$, however, it is always concave, with the tipping point between the alternating behaviours depending on the value of $\beta_t$.

\begin{figure}[t!]
	\centering{
        \includegraphics[height=0.2\textheight]{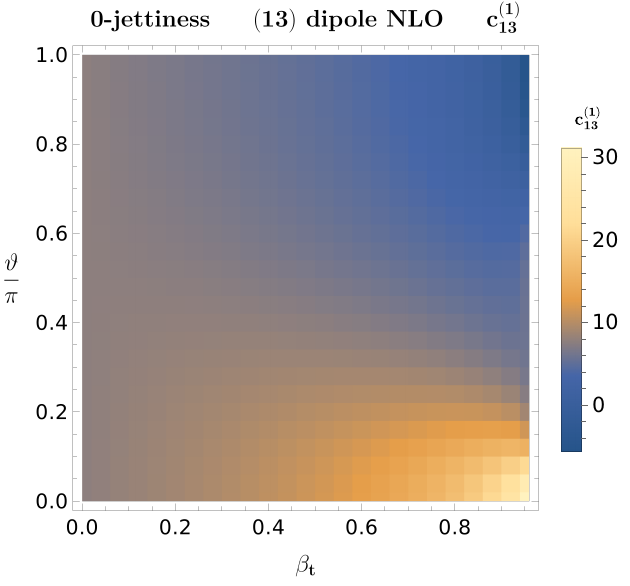}
        \includegraphics[height=0.2\textheight]{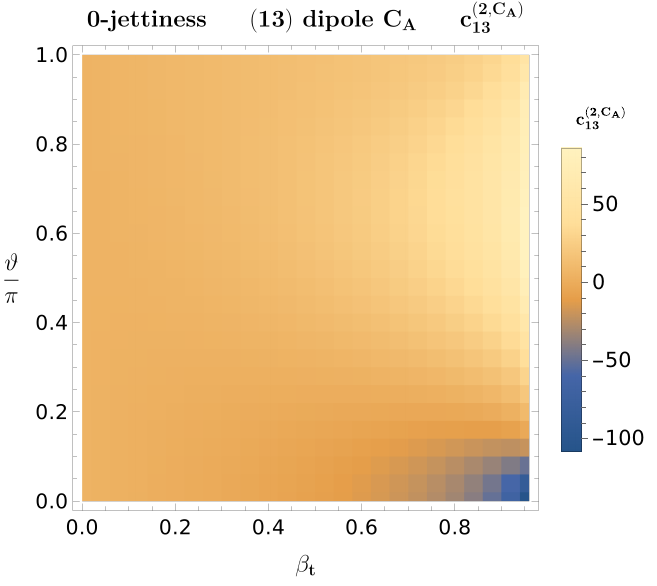}
        \includegraphics[height=0.2\textheight]{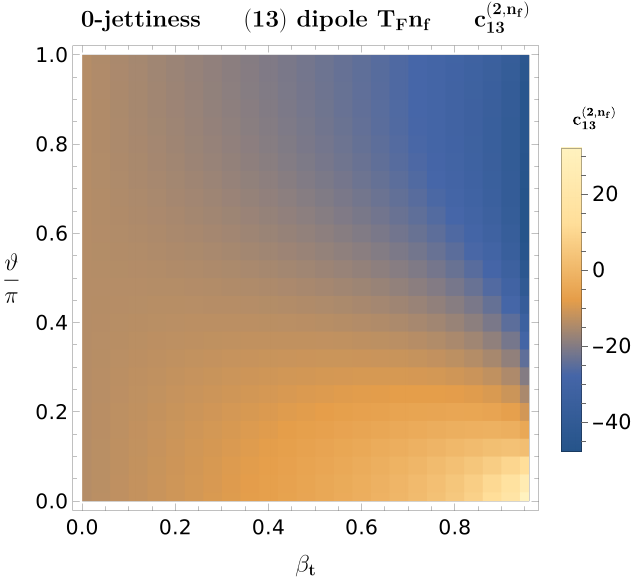}        
        \\[1.0em]
		\includegraphics[height=0.2\textheight]{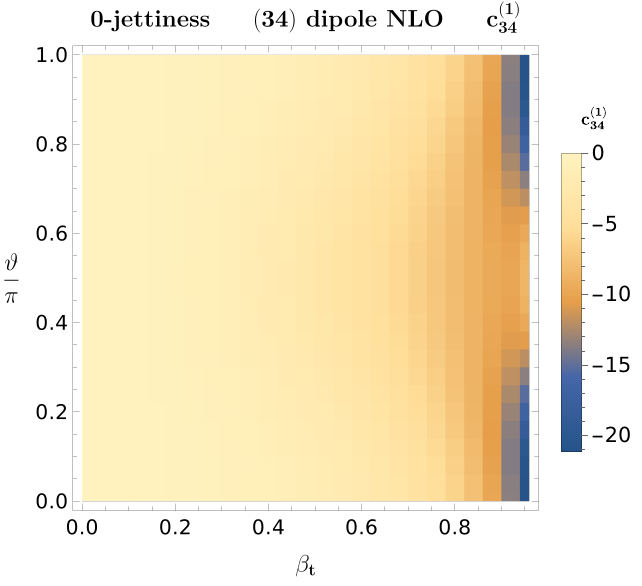}
        \includegraphics[height=0.2\textheight]{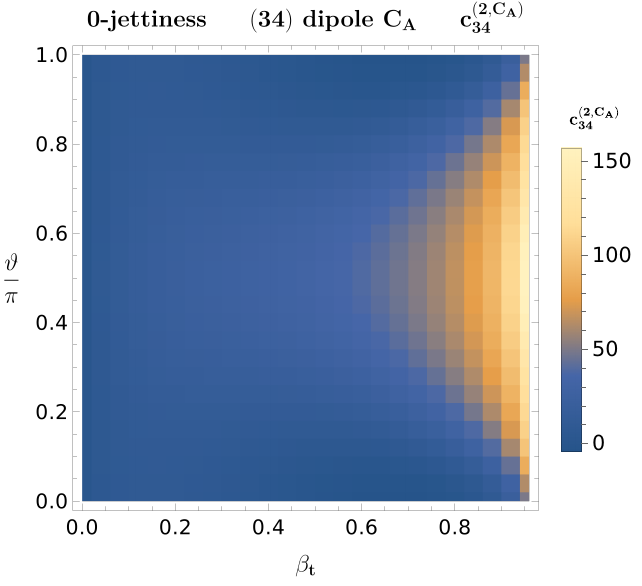}
        \includegraphics[height=0.2\textheight]{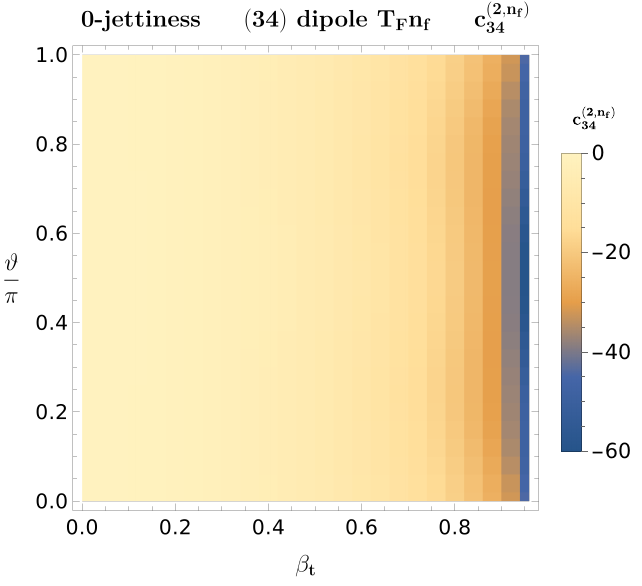}        
        \\[1.0em]
        \includegraphics[height=0.2\textheight]{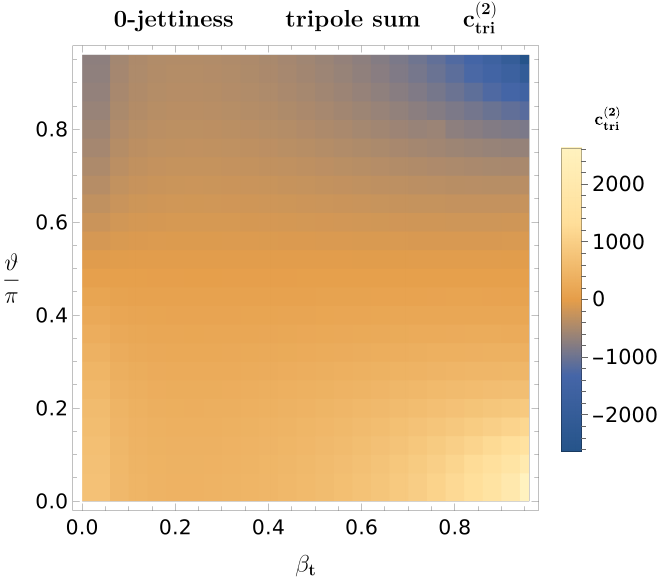}}
	\caption{Coefficients of the renormalised soft function~\eqref{eq:softrge} for the mixed (upper row) and purely massive dipole (middle row) at NLO (left column) and the two NNLO colour structures (middle and right column). The lower plot shows the renormalised tripole sum in the convention of \eqref{eq:c2tridef}.
    }
	\label{fig:fullgridresults}
\end{figure}

Finally, in Fig.~\ref{fig:fullgridresults} we show the full grid results for the renormalised coefficients in form of heatmaps, scanning over the full range of $\vartheta$ and $\beta_t$. Similar to the one-dimensional projections shown before, the plots display the raw data of our calculation without any type of interpolation between the pixels.

\section{Conclusions}
\label{sec:conclusions}

We have generalised the \texttt{SoftSERVE} framework, previously restricted to soft functions that arise in hard-scattering processes involving only massless partons, to processes with two final-state heavy quarks in back-to-back kinematics, as relevant for hadronic top-quark pair production. Specifically, we have focused on SCET$_\mathrm{I}$ observables that satisfy the non-Abelian exponentiation theorem in this work.

By employing suitable phase-space parameterisations in boosted dipole frames, we succeeded in keeping the required matrix elements in a compact form, absorbing the non-trivial angular dependence in the transverse space with respect to the dipole directions into the measurement function. This leads to noticeable improvements in the numerical evaluation of the phase-space integrals. As an independent cross-check, we also implemented a straightforward light-cone decomposition in the partonic centre-of-mass frame and found excellent numerical agreement between the two approaches.

We furthermore verified  numerically that the poles of the bare soft function agree with the prediction of the renormalisation-group equation. In section~\ref{sec:ren}, we presented a detailed derivation of all contributions entering the renormalisation of the soft function, with particular emphasis on the tripole colour correlations. These receive contributions both from commutators generated by the Baker-Campbell-Hausdorff expansion and from the genuine two-loop tripole term in the soft anomalous dimension. The former also induce a subtle finite contribution to the renormalised soft function, for which we derived an analytic expression at the end of that section.

We then applied the new framework to compute, for the first time, the NNLO \mbox{0-jettiness} soft function for hadronic top-quark pair production. In section~\ref{sec:results}, we illustrated several features of these results through representative one-dimensional projections, while the complete numerical results, including their uncertainties, are provided in the ancillary electronic files. We also derived the leading behaviour of the tripole contributions in the threshold limit analytically, which is required to reliably control the endpoint region. Since we kept the colour-generator structure fully generic throughout the derivation, without projecting onto a specific colour basis, a subset of our results are directly applicable to soft functions relevant for top-quark pair production at lepton colliders and single-top production at hadron colliders.

These results constitute a crucial ingredient for NNLL$^\prime$ resummation of 0-jettiness in top-quark pair production, as well as for the implementation of NNLO slicing and non-local subtraction methods, including Monte-Carlo event generators that match fixed-order calculations to parton showers. Natural extensions of the present work include the application of the framework to an even broader class of observables, as well as the computation of soft functions with heavy quarks in non-back-to-back kinematics. The latter are relevant for the associated production of a heavy-quark pair with a colour-singlet system and for processes involving a heavy-quark pair accompanied by additional massless partons. While the present formalism can be readily generalised to this class of soft functions, such an extension requires the introduction of additional angular variables, as discussed in the \texttt{SoftSERVE} extension for $N$-jet production with $N \geq 2$~\cite{Bell:2023yso}.\\

\noindent
\textbf{Note added:} We thank the authors of~\cite{Agarwal:2026dnj} for
pointing out a numerical artifact in our first set of numbers for the tripole contribution $c^{(2)}_\textrm{tri}$ that was due to a technical cutoff in our calculation. The issue was most pronounced for configurations with the top pair emitted orthogonal to the beam
directions, where it led to a deviation of a few permille that was not captured by our uncertainty estimate.
We also thank Rikkert Frederix for pointing out a relative bias of similar origin, at the $10^{-4}$ level at its most pronounced, in the finite NLO contributions $c^{(1)}_{13}$ and $c^{(1)}_{34}$.
The changes -- which are not visible in any of the plots -- are included in the electronic file accompanying the v2 of the paper.

\section*{Acknowledgments}

The authors would like to thank B.~Dehnadi for collaborating in the early stages of this project. A.B. would like to thank S.~Alioli and G.~Scala for discussions. The research of G.B.~and S.E.~was supported by the Deutsche Forschungs\-gemeinschaft (DFG, German Research Foundation) under grant 396021762 - TRR 257, and Germany’s Excellence Strategy – Cluster of Excellence ``Color meets Flavor'', EXC 3107 – Project-ID 533766364. M.A.L.~was supported by the UKRI guarantee scheme for the Marie Sk\l{}odowska-Curie postdoctoral fellowship, grant ref. EP/X021416/1. R.R.~was supported by the European Union’s Horizon Europe research and innovation programme under the Marie Sk\l{}odowska-Curie project ``SoftSERVE-NGL'' with grant agreement No. 101108359.
We thank the Erwin-Schr\"odinger International Institute for Mathematics and Physics at the University of Vienna for partial support during the Programme ``New Paradigms for Harnessing Quantum Field Theory at Colliders'', July 27 - August 28, 2026.

\appendix

\section{Anomalous dimensions}
\label{app:AD}

We collect in this appendix the relevant anomalous dimensions. The quantity
\begin{align}
X \in \{\beta, \Gamma, \gamma^{B,q},\gamma^{B,g},\gamma^q, \gamma^g, \gamma^Q, \Gamma(\beta), g(\beta) \} \, ,\nonumber
\end{align}
has the following expansion in terms of perturbative coefficients
\begin{align}
X(\alpha_s) = \sum_{n \geq 0}   \bigg(\frac{\alpha_s}{4 \pi}\bigg)^{n+1} \, X_n\, .
\end{align}
The QCD beta function up to two-loops is given by
\begin{align}
\beta_0 &= \frac{11}{3}C_A - \frac{4}{3}T_F n_f\,, \nonumber \\
\beta_1 &= \frac{34}{3}C_A^2 - \frac{20}{3}C_A T_F n_f - 4C_F T_F n_f \,,
\end{align}
where $T_F=1/2$ and $n_f$ is the number of active light quark flavours.
The coefficients of the cusp anomalous dimension are
\begin{align}
\Gamma_{0} &= 4\,,  \nonumber \\
\Gamma_{1} &= \left(\frac{268}{9} - \frac{4\pi^2}{3}\right)C_A - \frac{80}{9}\,T_F n_f\,.
\end{align}
The N-jettiness beam non-cusp anomalous-dimension coefficients for external quarks and anti-quarks are given by
\begin{align}
\gamma_0^{B, q} &= 6C_F\,, \nonumber \\
\gamma_1^{B, q} &= C_F\left[\left(\frac{146}{9} - 80\zeta_3\right)C_A 
+ \left(3 - 4\pi^2 + 48\zeta_3\right)C_F 
+ \left(\frac{121}{9} + \frac{2\pi^2}{3}\right)\beta_0\right]\,,
\end{align}
and the beam non-cusp anomalous-dimension coefficients for gluons are expressed as
\begin{align}
\gamma_0^{B,g} &= 2\beta_0\,, \nonumber \\
\gamma_1^{B,g} &= C_A\left[C_A\left(\frac{182}{9} - 32\zeta_3\right) 
+ \beta_0\left(\frac{94}{9} - \frac{2\pi^2}{3}\right)\right] + 2\beta_1\,.
\end{align}
The hard non-cusp anomalous dimensions for external quarks, anti-quarks $\gamma^q = \gamma^{\bar{q}}$ and gluons are given by
\begin{align}
\gamma_0^q &= -3C_F\,,  \nonumber \\
\gamma_1^q &= C_F^2\left(-\frac{3}{2} + 2\pi^2 - 24\zeta_3\right)
+ C_F C_A\left(-\frac{961}{54} - \frac{11\pi^2}{6} + 26\zeta_3\right)
+ C_F T_F n_f\left(\frac{130}{27} + \frac{2\pi^2}{3}\right)\,, \nonumber\\
\gamma_0^g &= - \beta_0 = -\frac{11}{3}C_A + \frac{4}{3}T_F n_f\,, \nonumber \\
\gamma_1^g &= C_A^2\left(-\frac{692}{27} + \frac{11\pi^2}{18} + 2\zeta_3\right)
+ C_A T_F n_f\left(\frac{256}{27} - \frac{2\pi^2}{9}\right)
+ 4C_F T_F n_f\,,
\end{align}
while the hard non-cusp anomalous-dimension coefficients for heavy quarks are given by
\begin{align}
\gamma_0^Q &= -2C_F\,, \nonumber \\
\gamma_1^Q &= C_F C_A\left(-\frac{98}{9} + \frac{2\pi^2}{3} - 4\zeta_3\right) + \frac{40}{9}C_F T_F n_f\,.
\end{align}
The cusp anomalous dimension for massive partons, which depends on the cusp angle $\beta_{IJ}$ has the following expansion coefficients
\begin{align}
\Gamma_{0}(\beta) &= \Gamma_{0}\,\beta\coth\beta\,, \nonumber \\
\Gamma_{1}(\beta) &= \Gamma_{1}\,\beta\coth\beta 
+ 8C_A\Bigg\{\frac{\pi^2}{6} + \zeta_3 + \beta^2 \nonumber\\
&\quad + \coth^2\beta\left[\mathrm{Li}_3(e^{-2\beta}) + \beta\,\mathrm{Li}_2(e^{-2\beta}) 
- \zeta_3 + \frac{\pi^2}{6}\beta + \frac{\beta^3}{3}\right] \nonumber\\
&\quad + \coth\beta\left[\mathrm{Li}_2(e^{-2\beta}) - 2\beta\ln(1-e^{-2\beta}) 
- \frac{\pi^2}{6}(1+\beta) - \beta^2 - \frac{\beta^3}{3}\right]\Bigg\}\,, 
\end{align}
and the function appearing in the genuine tripole contribution of the soft anomalous dimension is expressed as
\begin{align}\label{eq:gbeta}
g(\beta) &= \coth\beta\left[\beta^2 + 2\beta\ln(1-e^{-2\beta}) 
- \mathrm{Li}_2(e^{-2\beta}) + \frac{\pi^2}{6}\right] 
- \beta^2 - \frac{\pi^2}{6}\,.
\end{align}

\section{Colour matrices}
\label{app:CB}

In this appendix we provide explicit results for the matrix elements \eqref{eq:softmatrices} and \eqref{eq:softbornmatrix} in a colour basis that depends on the partonic channel. We adopt the singlet--octet basis defined as in~\cite{Ahrens:2010zv}:
\begin{align}\label{eq:basisvect}
    \left(c_1^{q\bar{q}}\right)_{\left\lbrace a\right\rbrace} &= \delta_{a_1 a_2}\delta_{a_3 a_4}, & \left(c_2^{q\bar{q}}\right)_{\left\lbrace a\right\rbrace} &= t^c_{a_2 a_1} t^c_{a_3 a_4},\nonumber \\
    \left(c_1^{\bar{q}q}\right)_{\left\lbrace a\right\rbrace} &= \delta_{a_2 a_1}\delta_{a_3 a_4}, & \left(c_2^{\bar{q}q}\right)_{\left\lbrace a\right\rbrace} &= t^c_{a_1 a_2} t^c_{a_3 a_4},\nonumber \\
    \left(c_1^{gg}\right)_{\left\lbrace a\right\rbrace} &= \delta^{a_1 a_2}\delta_{a_3 a_4}, &\left(c_2^{gg}\right)_{\left\lbrace a\right\rbrace} &= i f^{a_1 a_2 c} t^c_{a_3 a_4}, & \left(c_3^{gg}\right)_{\left\lbrace a\right\rbrace} &= d^{a_1 a_2 c} t^c_{a_3 a_4}.
\end{align}
For the $q\bar q$ channel ($\dR=N$) we find
\begin{align}
 \mathbf{w}^{q \bar{q}}_{\mathrm{Born}}&=\mat{N&0\\[2pt]0&\dfrac{\CF}{2}}\, ,\nonumber \\
 \w^{q \bar{q}}_{12}=\w^{q \bar{q}}_{34}&=-\frac{\CF}{4N}\mat{4N^2&0\\0&-1}\, ,\nonumber\\
 \w^{q \bar{q}}_{13}=\w^{q \bar{q}}_{24}&=-\frac{\CF}{2}\mat{0&1\\[2pt] 1 \;\;\;& 2\CF -\frac{N}{2}} \, ,\nonumber\\
 \w^{q \bar{q}}_{14}=\w^{q \bar{q}}_{23}&=-\frac{\CF}{2N}\mat{0&-N\\-N&1}\, ,\nonumber\\
\w^{q \bar{q}}_{\tr} &=\frac{N\CF}{4}\mat{0&-i\\ i&0}=\frac{N^2-1}{8}\mat{0&-i\\ i&0} \, .
\end{align}
For the $\bar q q$ channel we find instead the following relations $\w^{ \bar q q}_{23}=\w^{q \bar q}_{13}$, $\w^{\bar q q}_{13}=\w^{q \bar q}_{23}$,
$\w^{ \bar q q}_{24}=\w^{q \bar q}_{14}$, $\w^{\bar q q}_{14}=\w^{q \bar q}_{24}$, $\w^{\bar{q} q}_{\tr}=-\w^{q \bar{q}}_{\tr}$. For the $gg$ channel ($\dR=N^2-1$) we find
\begin{align}
\mathbf{w}^{gg}_{\mathrm{Born}}&=\mat{N&0&0\\[2pt]0&\dfrac N2&0\\[2pt]0&0&\dfrac{N^2-4}{2N}}\, ,\nonumber\\
\w^{gg}_{12}&=-\frac14\mat{4N^2&0&0\\0&N^2&0\\0&0&N^2-4} \, ,\nonumber\\
\w^{gg}_{34}&=-\mat{N\CF&0&0\\[2pt]0&-\dfrac14&0\\[2pt]0&0&-\dfrac{N^2-4}{4N^2}}\, , \nonumber\\
\w^{gg}_{13}=\w^{gg}_{24}&=-\frac18\mat{0&4N&0\\4N&N^2&N^2-4\\0&N^2-4&N^2-4} \, ,\nonumber
\end{align}
\begin{align}
\w^{gg}_{14}=\w^{gg}_{23}&=-\frac18\mat{0&-4N&0\\-4N&N^2&-(N^2-4)\\0&-(N^2-4)&N^2-4} \, ,\nonumber\\
\w^{gg}_{\tr} &=\frac{N^2}{4}\mat{0&-i&0\\ i&0&0\\0&0&0} \, .
\end{align}

We would like to stress that the matrix representations of operators in colour space $[\mathbf{O}]_{MN}$ in an orthogonal (but not orthonormal) basis can be extracted by rescaling the matrix elements with the appropriate normalisation factor for the basis vectors such that
\begin{align}
    [\mathbf{O}]_{MN} = \frac{\langle c_M |\mathbf{O} | c_N\rangle}{\langle c_M | c_M \rangle}\, ,
\end{align}
where $\mathbf{O}$ represents in this context a dipole or tripole colour operator. This is, for instance, the form in which the anomalous dimensions enter the RG equations in a given basis.

\bibliographystyle{jhep}
\bibliography{bibliography}
\end{document}